\documentclass[a4paper,11pt]{article}
\pdfoutput=1 

\usepackage{jcappub} 

\usepackage[T1]{fontenc} 
\usepackage{comment}
\usepackage{xspace}

\usepackage{amsmath}
\usepackage{amssymb}
\usepackage{rotating}
\usepackage{lineno}
\usepackage{multirow}
\usepackage{colortbl}
\usepackage{xcolor}
\graphicspath{{figs/}}

\newcommand{\lsim}{\raisebox{-0.13cm}{~\shortstack{$<$ \\[-0.07cm]
      $\sim$}}~}
      
\newcommand{\beq}{\begin{equation}}
\newcommand{\eeq}{\end{equation}}

\newcommand{\nmu}{N_\mu}

\newcommand{\Xmumax}{X^{\mu}_{\rm max}}

\def\sib{Sibyll\,2.3d\xspace}
\def\sibOld{Sibyll\,2.1\xspace}
\def\qgs{QGSJet\,II-04\xspace}
\def\qgsOld{QGSJet\,II-03\xspace}
\def\epos{EPOS-LHC\xspace}
\def\eposOld{EPOS~1.99\xspace}

\title{\boldmath Universality of the muon component of extensive air showers}

\author[a]{L. Cazon}
\author[b,c]{R. Concei\c{c}\~{a}o}
\author[a,b]{F. Riehn}

\affiliation[a]{Instituto Galego de F\'isica de Altas Enerx\'ias (IGFAE), University of Santiago de Compostela, Rúa de Xoaquín Díaz de Rábago, Santiago de Compostela, Spain}
\affiliation[b]{Laborat\'orio de Instrumenta\c{c}\~ao e F\'isica Experimental de Part\'iculas (LIP), Avenida Prof. Gama Pinto, Lisbon, Portugal}
\affiliation[c]{Departamento de F\'isica, Instituto Superior T\'ecnico, Av. Rovisco Pais, Lisbon, Portugal}

\emailAdd{lorenzo.cazon@usc.es}
\emailAdd{ruben@lip.pt}
\emailAdd{friehn@lip.pt}

\abstract{
In extensive air shower experiments, the number of muons crossing a detector at a given position, as well as their arrival time, arrival direction, and energy, are determined by a more fundamental 3-dimensional distribution linked to the hadronic core of the shower. Muons are produced high up in the atmosphere after the decay of mesons in the hadronic cascade. The distributions of production depth, energy, and transverse momentum of muons are enough to fully predict the muon component of air showers in any particular observational condition.  By using air-shower simulations with the state-of-the-art hadronic interaction models, the mentioned distributions at production are analyzed as a function of zenith angle, primary mass, and hadronic interaction model, and their level of universality is studied and assessed in an exhaustive manner for the first time.
}

\begin{document}
\maketitle
\flushbottom

\section{Introduction}
\label{s:int}

Ultra-High Energy Cosmic Rays (UHECRs) are particles continuously entering the Earth's atmosphere. They are of interest for two reasons: on the one hand, they are produced in outer space, bringing us valuable information about their astrophysical sources and the interstellar medium; on the other hand, they can reach energies several orders of magnitude beyond those attained in human-made accelerators, allowing us to peek into the physics at energies beyond the Large Hadron Collider (LHC).

When a UHECR enters the atmosphere, it collides with an air nucleus producing secondary particles, which keep interacting in successive reactions producing even more particles, creating a so-called Extensive Air Shower of particles (EAS). Many of the main properties of EAS such as the number of particles and their energy distribution are the same regardless of the type UHECR that initiates the shower. This has been commonly known in literature as {\it shower universality}. 

The composition of UHECRs can be derived by comparing certain shower observables (like the depth of maximum of the electromagnetic shower, $X_{\rm max}$) with the corresponding simulated ones for different primary composition scenarios. Given that EAS particle reactions occur at energies and in phase-space regions out of the reach of Earth-based accelerators, our current understanding of EASs is subject to significant theoretical uncertainties, which directly translate into uncertainties in the simulation of the EAS observables, and, therefore, in the final interpretation of primary-mass composition. On the other hand, some other EAS properties might be unaffected by uncertainties in hadronic interactions. This allows to expand the concept of {\it shower universality} against certain changes of the hadronic interactions.

An EAS can be separated into two major components: the hadronic and the electromagnetic (EM) component (also known as the hadronic and the EM cascade). The EM component consists of photons and electrons and positrons. The EM cascade is fed from the hadronic cascade by the decay of neutral pions $\pi^0$ into photons, which then undergo pair production and Bremsstrahlung.
The electromagnetic cascade is characterized by several important features which have been extensively studied in the past. The bulk of electromagnetic particles exhibits a universal energy spectrum that depends only on shower age, and the angular distribution of electrons depends only on their energy~\cite{Giller:2004cf,Nerling:2005fj,Lipari:2009zz,Lafebre:2009en,Giller:2014uja,Smialkowski:2018reh}. The longitudinal development can be expressed by a universal Gaisser-Hillas function, whose parameters are independent of the mass of the primary, and only the energy and depth of the first interaction of the UHECR are enough to determine all characteristics of the EM shower. Naturally, this statement is only valid to a certain degree of detail, where violation of universality starts to be visible (see for instance Ref.~\cite{Lafebre:2009en,USP1,USP2}). 

The hadronic component is comprised of mesons (mainly charged pions, since neutral pions rapidly decay into photons) and baryons. As the hadrons interact and create new particles, the average energy per hadron decreases. At a certain point, it becomes more likely that mesons decay rather than interact. Muons, a common product of meson decays, lose very little energy in their propagation in the atmosphere and thus trace the development of the hadronic cascade. 

Similar to its EM counterpart, the muonic component also presents universality features: the spatial distribution of muons arriving at the ground does not depend on the type of the primary (its mass) and its energy~\cite{Blake:1995ft,Ave:2000xs}. This universality of the shower properties is what allows experiments to reconstruct the muon component in each shower independently of the knowledge of the mass of the primary. 

Beyond the so-called {\it pure muon component} (directly emerging from the hadronic cascade via $\pi^\pm$ decay), and the {\it pure EM component}, (from high energy $\pi^0$ decays), other contributions were identified~\cite{Schmidt:2007vq,Ave:2017uiv,Ave:2017wjm} which must be accounted for, namely:  {\it  EM from muon decay} or {\it muon halo} which stems from the decay of muons, and therefore scales with the hadronic component of the shower;
  {\it  EM from low-energy $\pi^0$ decay} which is a small contribution to the EM cascade but nevertheless is coupled with the hadronic cascade;
  and {\it  muon from photo-production}, which stems from the pion production after photon-air interactions, and is therefore coupled to the EM cascade.
  
The number of muons at the ground as a function of the primary energy has been measured in the Pierre Auger Observatory~\cite{Aab:2014pza,PierreAuger:2021qsd} using inclined showers, where the EM component is fully absorbed in the atmosphere and only muons reach the observation level. An excess at the level of $26$ to $43\%$ with respect to predictions was found.
In Ref.~\cite{Aab:2016hkv}, using vertical showers detected at the Pierre Auger Observatory and assuming a mixed primary composition, the hadronic scaling factor was measured to be betwen $1.33$ and $1.45$. In the same measurement, it was found that the EM component, responsible for the overall energy scale, agrees with expectations. 
A comprehensive study of the muon number measurements done by different experiments~\cite{WHISP} points to a increase of the muon number with energy relative to simulations over a wide range of energies. This discrepancy in the muon component between observation and expectation is commonly referred to as the ``muon puzzle'' in EAS. 
One of the difficulties of the compilation of measurements of the muon content of EASs in different experiments is to account for the particular observation conditions of each experiment. The study of the muon distributions in literature has been performed at the ground, or more generally, at the detector level. It is important to notice that propagation of muons from the production point to the detector introduces a set of non-trivial effects in the resulting energy, production depth, arrival time, and lateral distribution~\cite{Cazon:2012ti} which depend on the particular observation conditions, like the distance to the shower core, energy threshold, slant depth of the muon detectors, and zenith angle of the shower. However, at the moment of production (charged meson decay), the distributions of muons strictly follow the development of the hadronic cascade. Therefore, the universal features of the hadronic cascade are much better traced by the muon distributions at production, the muon propagation being a problem that can be treated and understood separately. Moving our focus from the distributions of muons at ground level to the distribution of muons at production greatly simplifies the problem of understanding and assessing the universality features, eventually allowing for more precise and manageable descriptions with less independent parameters. 

This paper assesses for the first time the degree of universality of the distributions of muons at production for the relevant variables by using simulations.
  
The air-shower simulations used in this work were done using CORSIKA~v7.7402 Monte Carlo tool~\cite{Heck98a}. Hadronic interactions of particles with an energy below $80\,$GeV are simulated with FLUKA~v2011.2c~\cite{fluka,fluka2} and URQMD~\cite{Bleicher99a}. 
For hadronic interactions above this threshold we use the current post-LHC models (\epos~\cite{Pierog:2009zt,Pierog:2013ria}, \sib~\cite{sib23d,Ahn:2009wx}, and \qgs~\cite{Ostapchenko:2010vb,Ostapchenko:2013pia}), as well 
as the interaction models that predate the LHC measurements (\eposOld~\cite{Pierog:2009zt}, \sibOld~\cite{Ahn:2009wx}, and \qgsOld~\cite{Ostapchenko:2010vb}) to get a more complete picture of the physically allowed phase-space of air showers. We obtain the distributions at production by filling histograms during the shower simulations using an adjusted COAST interface~\cite{Heck98a} (based on ROOT~\cite{Brun:1997pa}). We simulate $100$ showers for proton and iron primaries with a primary energy of $10^{19}\,$eV and zenith angles of $0^\circ$, $20^\circ$, $40^\circ$, $60^\circ$ and $70^\circ$. Higher zenith angles are not explored to avoid changing the air shower simulation to a curved atmosphere. To improve the runtime of the shower simulations CORSIKA is used with the thinning option~\cite{Heck:1998gr,Hillas:1997tf}. The thinning algorithm is configured to group together statistically all particles below an energy of $10\,$TeV with a maximal weight of $10^4$.
For the study of individual hadronic interactions we used the IMPY framework which provides a standardized interface to the interaction models mentioned above~\cite{impy}.
The figures were made with Matplotlib~\cite{Hunter:2007}.

\section{The muon component}

After the first UHECR-air interaction, approximately $\sim 75\%$ of the energy goes into secondary mesons (excluding neutral pions) and baryons~\cite{Cazon:2019mtd}, which continue interacting, creating the so-called {\it hadronic cascade}. When the average energy per meson decreases, mesons eventually become more likely to decay rather than interact. This energy is called the critical energy, and it marks the stage of the shower where most muons are formed. 

Provided that $({\mathbf x}_\text{i},{\mathbf p}_\text{i})$ is the position and momentum of the muon at the production point, and $({\mathbf x}_\text{f},{\mathbf p}_\text{f})$ is the position and momentum of the muon when arriving at a given detector surface, located for instance at the ground, then, the muon distributions at production and at the detector are related through a transformation:
\beq
\frac{\mathrm{d}^6 N }{\mathrm{d} {\mathbf x}_{\rm i} \mathrm{d} {\mathbf p}_{\rm i}} \,\,\stackrel{\text{propagation}}{\longrightarrow} \,\, \frac{\mathrm{d}^6 N }{\mathrm{d} {\mathbf x}_{\rm f} \mathrm{d} {\mathbf p}_\text{f}}
\eeq
where the $\stackrel{\text{propagation}}{\longrightarrow}$ stands for all interactions undergone by muons in their propagation through the atmosphere, mainly energy loss, decay, scattering, deflection due to the geomagnetic field, and Bremsstrahlung.
 Due to the azimuthal symmetry of muon production around the shower axis, the number of dimensions necessary to describe muons at production can be reduced~\cite{Cazon:2012ti}. In addition, because of the kinematics of the parent mesons, most muons are produced within a few tens of meters from the shower axis~\cite{Cazon:2012ti,Cazon:2004zx}. As such, we can make the approximation that all muons are being produced on the shower axis. Within this approximation, the position and the 3-momentum at production can be reduced to the height $(z)$ and $(p_{\rm T},p_z)$. More conveniently, these are expressed in terms of: production slant depth $X$, accounting for the traversed matter along the $z$-axis, the total energy $E_\text{i}$, and the transverse momentum $p_{\rm T}$ with respect to the shower axis.

The function fully describing the muon distributions at production is therefore,
\beq
\frac{\mathrm{d}^{3} N}{\mathrm{d}X\, \mathrm{d}E_\text{i}\,\mathrm{d}p_{\rm T}}= F(X,E_\text{i},p_{\rm T}).
\label{eq:mu}
\eeq 
To calculate the distributions at the ground or any other observational conditions, muons should be propagated as explained in Ref.~\cite{Cazon:2012ti}.

The muon production depth (MPD) distribution is simply
\begin{equation}
  \frac{\mathrm{d}N}{\mathrm{d}X} \equiv h(X)= \iint  F (X,E_\text{i},p_{\rm T})\, \mathrm{d}E_\text{i} \, \mathrm{d}p_{\rm T} \, .
\end{equation}
This distribution displays a maximum at ${\cal \Xmumax}$. We can therefore rewrite the equation and have the distributions {\it centered} around ${\cal \Xmumax}$:
\beq
\frac{\mathrm{d}^{3} N}{\mathrm{d}X'\, \mathrm{d}E_\text{i}\,\mathrm{d}p_{\rm T}}={\cal \nmu} \, f(X',E_\text{i},p_{\rm T}) \ ,
\label{eq:mu1}
\eeq
where
\begin{equation}
  f(X',E_\text{i},p_{\rm T})\equiv \frac { F({\cal \Xmumax}+X',E_\text{i},p_{\rm T})}{\cal \nmu} \ , \nonumber
\end{equation}
and $X'=X-{\cal \Xmumax}$ and ${\cal N}_\mu=\int h(X) \, \mathrm{d}X = \iiint F(X,E_\text{i},p_{\rm T})\,\mathrm{d}X \, \mathrm{d}E_\text{i} \, \mathrm{d}p_{\rm T}$ is the total number of muons produced in the shower. It is worth noting the difference between $h(X)$, called {\it total/true} MPD, which counts the production rate of all muons along the shower axis, with respect to the {\it apparent} or {\it propagated} MPD, defined as $\mathrm{d}^4N/\mathrm{d} {\mathbf x}^3_\text{f} \mathrm{d}X(X)$, which counts the muons produced in a given interval $\mathrm{d}X$ and which are arriving at a given interval $\mathrm{d} {\mathbf x}^3_\text{f}$. The {\it apparent} MPD is different from $h(X)$ because it contains only those surviving muons which were emitted from the shower axis with the appropriate solid angle, thus restricting the $E, p_{\rm T}$ phase-space. The {\it apparent} MPD and its corresponding depth of maximum $\Xmumax$, is the one reconstructed by experiments, like the Pierre Auger Observatory~\cite{Aab:2014dua}.

\begin{figure}
  \centering
  \includegraphics[width=0.85\textwidth]{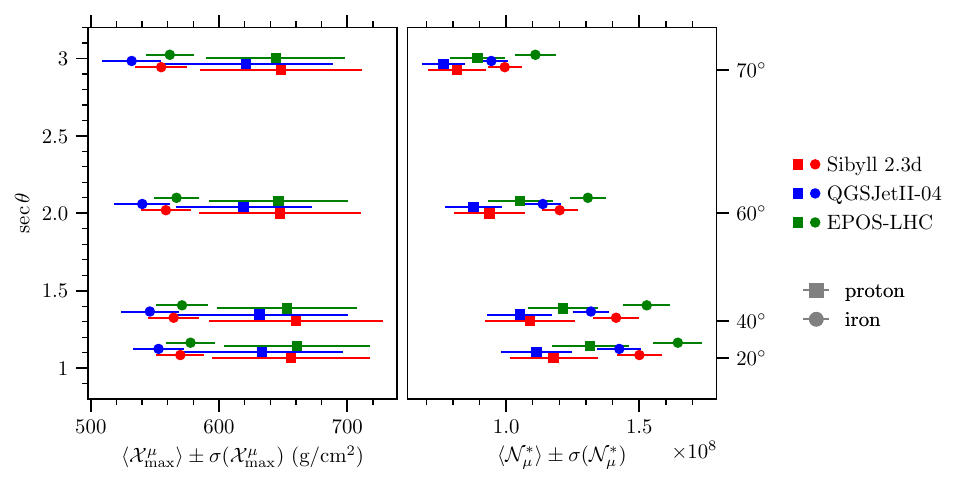}
  \caption{Evolution of the mean of the $\mathcal{X}^\mu_{\rm max}$ (left panel) and $\mathcal{N}^*_\mu$ (right panel) distributions with zenith angle and primary particle type. The width of the distributions is represented by the error bars. Note that all showers were calculated for zenith angles of $20^\circ$, $40^\circ$, $60^\circ$ or $70^\circ$. To enhance readability different hadronic interaction models were shifted in zenith angle.
    \label{fig:xmumax-nmu-average}}
\end{figure}

In Figure~\ref{fig:xmumax-nmu-average} we show the behaviour of the normalizations of Eq.~\eqref{eq:mu1}, the average value of the depth of maximum production ($\mathcal{X}_{\rm max}^\mu$) and the total number of produced muons (${\cal N}_\mu$), for different primaries and zenith angles. The average depth of maximum production, $\langle \mathcal{X}^\mu_{\rm max} \rangle$, changes by $\approx 100\,$g$/$cm$^2$ between proton and iron showers but only by at most $20\,$g$/$cm$^2$ between the different zenith angles. The mean value variation of $\mathcal{X}^\mu_{\rm max}$ between the different interaction models is $27\,$g$/$cm$^2$. Much like the maximum of the electromagnetic shower profile ($X_{\rm max}$)~\cite{KampertUnger,Aab:2014kda} the distributions of proton and iron primaries are not that well separated when taking into account the widths of the distributions (indicated by the errorbars in the figure). For the number of muons, instead of ${\cal N}_\mu$ we have calculated ${\cal N}_\mu^*=\int_{-\infty}^{X'_{\rm up}} h(X')\, \mathrm{d}X'$ where the upper limit is set to $X'_{\rm up}={100\,\mathrm{g}/\mathrm{cm^2}}$, that is $100\,$g$/$cm$^2$ after the maximum of production, to avoid the truncation of the MPD distribution, $h(X)$, due to the ground level. In vertical showers, the maximum of the MPD is only around $\sim300\,$g$/$cm$^2$ above the ground at the Auger site, impeding $\sim20\%$ of muons with large production depth from being produced. For zenith angles beyond $40^\circ$, $\langle \mathcal{N}_\mu^{*} \rangle$ moderately decreases with increasing zenith angle, to such an extent that an iron induced shower at $70^\circ$ produces less muons than a vertical proton shower at the same energy of the primary. Note that our definition of $\langle \mathcal{N}_\mu^{*} \rangle$ is such that the accounting of muons is done at the same slant depth ($100\,\mathrm{g}/\mathrm{cm^2}$ after the maximum). Therefore modifications in the total number of muons come only from changes in the critical energy of hadrons due to the development of the showers in regions with different atmospheric densities.

In Ref.~\cite{Andringa:2011ik}, it was observed that muons at production display universal features when at the same stage of shower development measured of $X'$. In this paper, we assess the degree of universality of the relevant muon distributions, which can be reduced to the 3-dimensional normalized function $f(X',E_\text{i},p_{\rm T})$ and its projections in $X'$, $p_{\rm T}$ and $E_{i}$.

\section{Transverse momentum distribution}
\label{sec:pt}

\subsection{Transverse momentum of muons, hadron decay, hadron production}
\vspace{0.1cm}
Muons mainly come from the decay of pions and kaons, either directly or through the decay of intermediate mesons (kaons will decay with a 28.5\% probability into charged pions again).

For muons in a 2-body decay, which is the main decay mode of both pions and kaons, simple kinematics show that on average, the transverse momentum of the muon with respect to the trajectory of the parent meson is $21\,$MeV in the case of pions and $167\,$MeV in the case of kaons.

\begin{figure}
  \centering
  \includegraphics[width=0.45\textwidth]{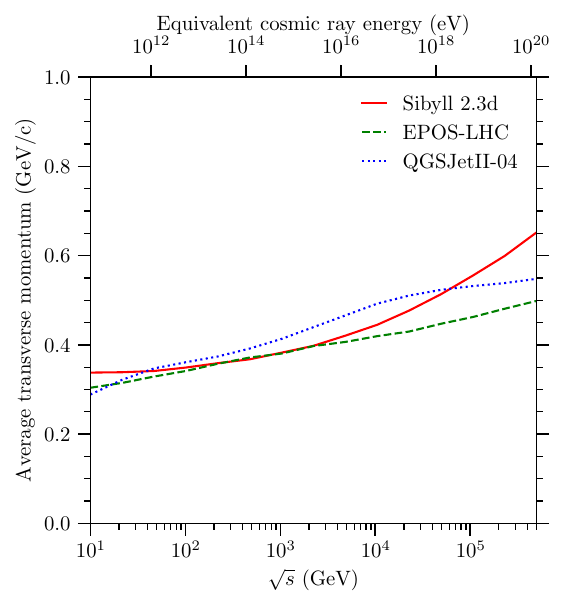}
  \caption{The average transverse momentum of charged pions as a function of the center-of-mass energy $\sqrt{s}$ in proton-nitrogen interactions.
    \label{fig:avg-pt}}
\end{figure}

In a hadronic collision most of the produced hadrons are pions followed by kaons ($\approx 3$ times less frequent). Apart from pions and kaons, leading baryons can carry a large fraction of the energy producing similar reactions. The experimental data~\cite{Wong:1995jf,STAR:2003oii,Sirunyan:2017zmn} of such reactions  -- available up to a few tens of TeV per nucleon in the center-of-mass frame --  show a transverse momentum $\bf{p_{\rm T}}$ distribution that decreases exponentially,
\beq
    {\mathrm{d} N \over  \mathrm{d} {\bf p_{\rm T}}}={\mathrm{d} N \over 2\pi p_{\rm T} \mathrm{d} p_{\rm T}}  \propto \exp \left (- {p_{\rm T} \over Q} \right ) \ .
    \label{eq:hadron-pt}
\eeq
The slope $Q$, which is related to the average $p_{\rm T}$, changes slowly with the energy of the collision and the rapidity region. The average transverse momentum is of the order of tenths of GeV$/$c (see Fig.~\ref{fig:avg-pt}). On the other hand, the additional transverse momentum of muons from the pion and kaon decay mix ($\sim 0.07\,$GeV$/$c) is approximately 20\% of the total $p_{\rm T}$. The distribution of the outgoing muons can therefore be considered to be dominated by that of the parent hadrons. This is a critical feature responsible for many of the observed characteristics of the hadronic and muonic showers, as will be seen in further sections.

The origin of the transverse momentum of the parent hadrons of the muons (pions or kaons) is in the structure of the initially colliding hadrons. In the reference frame where the $z$-axis is aligned with the parent hadron, momentum space is again essentially two-dimensional: transverse and longitudinal. Prior to any interaction, the transverse momentum of the hadron is, by construction, zero. However, seeing hadrons as composite objects that are constituted by quarks and held together by gluons, even prior to any interaction, the constituents do carry transverse momentum due to Fermi motion. This transverse momentum, also called Fermi momentum, is
\begin{equation}
  p_{\rm T} \sim \hbar c/R_{\rm had} = 0.197\,\mathrm{GeV} \left ( {1\,\mathrm{fm} \over R_{\rm H}} \right ) \ ,
\end{equation}
where $R_{\rm H}=0.841\,$fm, $0.659\,$fm, or $0.560\,$fm for protons, pions, and kaons respectively~\cite{ParticleDataGroup:2020ssz}. When quarks and gluons scatter off one another in a hadronic interaction, they receive a \emph{kick}, that is, additional transverse momentum. Due to quantum fluctuations, the structure of hadrons is not fixed but changes from collision to collision. In particular, as the collision energy increases and the momentum that is in principle available for transfer increases, more and more short-lived fluctuations can play a role. This leads to the rise in the number of elementary scatterings between quarks and gluons (as reflected in the hadron multiplicity and cross section) and the growth of the average transverse momentum with energy, as can be seen in Figure~\ref{fig:avg-pt}. A direct comparison of the Fermi momentum with the observed average $p_{\rm T}$ reveals that scattering contributes up to 50\% of the total transverse momentum.

Pions and kaons in EASs typically interact several times ($\mathcal{O}(10)$) before they decay and produce muons. The final transverse momentum of these cascading mesons is determined by the $p_{\rm T}$ of the last interaction. In each interaction $j$, the meson emerges under an angle $\alpha_j \simeq p_{\mathrm{T} j}/E_j$ (where $E_j$ and $p_{{\rm T}j}$ are the energy and transverse momentum of the emerging meson) with respect to the direction of the parent particle, and a random azimuthal angle $\phi$. By fixing $\phi$ for all reactions in a given branch, that is,  all interactions happen in the same  plane, one can calculate the upper bound for the angle with respect to the shower axis for a cascading particle after $i$ interactions: $\theta^{\rm max}_i = \sum_{j=1}^i \alpha_j$. The final transverse momentum with regard to the shower axis, $p^{\rm sh}_{\mathrm{T}i}$, becomes
\begin{equation}
    p^{\rm sh}_{{\rm T}i} \lsim \theta^{\rm max}_i \, E_i = \sum_{j=1}^i p_{{\rm T}j} \frac{E_i}{E_j} \ ,
\end{equation}
which is the sum of transverse momenta gained in each interaction but weighted by the ratio of the energy at interaction $i$ over the energy in the interaction $j$. For $i=j$ the weight is 1 and it rapidly decreases as $j$ decreases\footnote{If in the interaction $j$, the secondary particle takes a fraction $x_j$ of the energy of the preceeding particle $E_{j-1}$ such that $E_j= x_j E_{j-1}$ , then we have that $E_i/E_j=\prod_{k=j+1}^i x_j$, which is a number that exponentially decreases as the difference $i-j$ increases. }.
The final momentum of the muons is mostly determined by the transverse momentum obtained in the last interaction $i$ that produced the decaying pion (meson),
\begin{equation}
  p^{\rm sh}_{{\rm T}i} \simeq  p_{{\rm T}i} \ .
\end{equation}

\subsection{Universality of the transverse momentum spectra}
\vspace{0.1cm}

\begin{figure}
  \centering
  \includegraphics[width=0.45\textwidth]{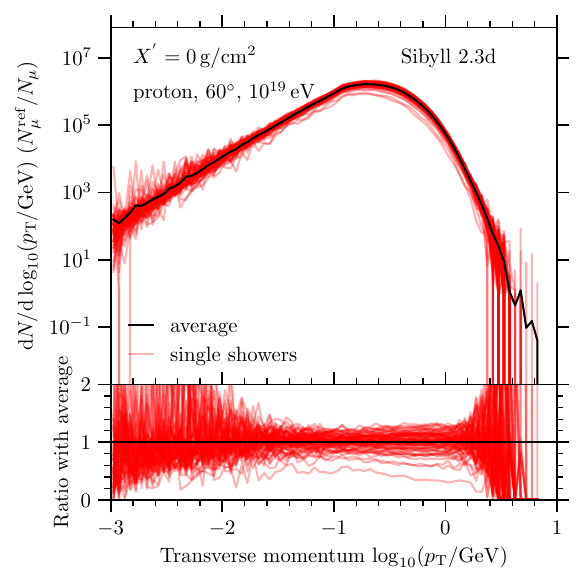}
  \qquad
  \includegraphics[width=0.45\textwidth]{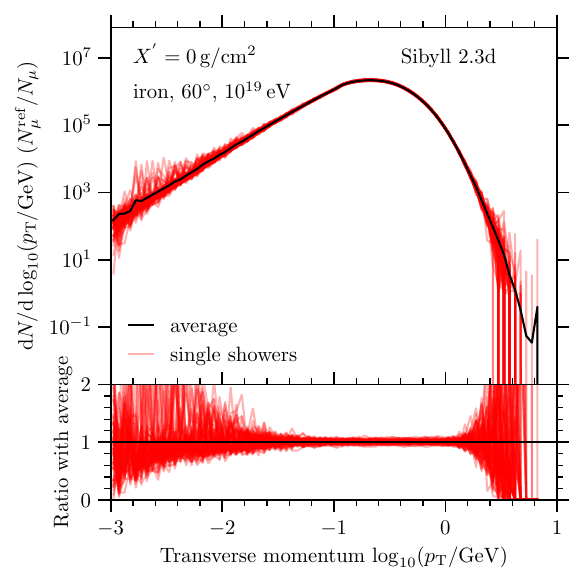}
  \caption{Transverse-momentum spectrum of muons at the maximum of muon production ($X^\prime=0$) for proton (left) and iron (right) induced air showers. The primary energy is $10^{19}\,$eV and the zenith angle is $60^{\circ}$. The spectra for 100 individual showers are shown in red. The average spectrum is shown in black. The top of each figure shows the differential spectrum, the bottom shows the ratio of the spectra with the average spectrum.  \label{fig:pt-spec-indiv}}
\end{figure}

Examples of the spectra of transverse momentum of muons in extensive air showers are shown in Figure~\ref{fig:pt-spec-indiv}. In the figure the individual $p_{\rm T}$-spectra of an ensemble of 100 air showers are shown together with the average spectrum. The spectra were taken at the maximum of muon production ($X^{'}=0$), the primary energy is $10^{19}\,$eV, the shower axis zenith angle $\theta = 60^\circ$, and the high-energy interaction model is \sib. In the left panel proton primaries and in the right panel iron primaries are shown. To reveal differences in shape both spectra are normalized to the number of muons in proton showers. At the bottom of each panel the ratio between the $p_{\rm T}$-spectrum of each individual shower with the average $p_{\rm T}$-spectrum is shown. As indicated in the previous section the bulk of the $p_{\rm T}$-spectrum ($-1.5<\log_{10}(p_{\rm T}/\mathrm{GeV})<0.1$) shows a high degree of universality both between individual showers of the same primary and between primaries (the scales in both panels are identical). The fluctuations at low transverse momenta are presumably artificial fluctuations introduced by the thinning algorithm, while the high-$p_{\rm T}$ fluctuations are from undersampling as there are very few muons per shower with a large $p_{\rm T}$.

\begin{figure}
  \begin{center}
    \includegraphics[width=0.45\textwidth]{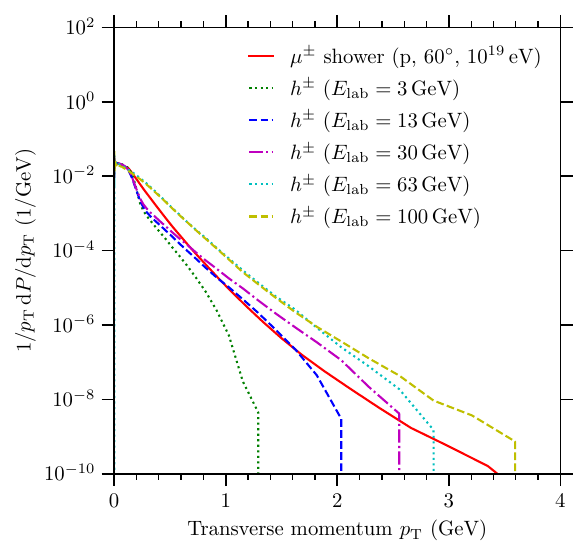}
   \caption{The average $p_{\rm T}$-spectrum of muons for proton showers compared with the $p_{\rm T}$-spectrum of charged hadrons in single p--N interactions at different energies. The distributions below lab.\ energy of $63\,$GeV were calculated with URQMD~\cite{Bleicher99a,impy}.}
   \label{fig:pt-spec-mu-vs-had}
  \end{center}
\end{figure}

The reason for this high degree of universality is that the transverse momentum of the muons is predominately due to the transverse momentum that mesons obtain in their last interaction before decaying to produce muons. In Figure~\ref{fig:pt-spec-mu-vs-had}, the distribution of $p_{\rm T}$ of muons in EASs and charged hadrons in single hadronic interactions for different laboratory energy are compared. The closest match for the bulk of the $p_{\rm T}$-distribution of muons are the distributions in the energy range from $3\,$GeV to $30\,$GeV since those are the typical energies in the last interaction before the muon is produced (the critical energy). 

\begin{figure}
   \centering 
      \includegraphics[width=0.45\textwidth]{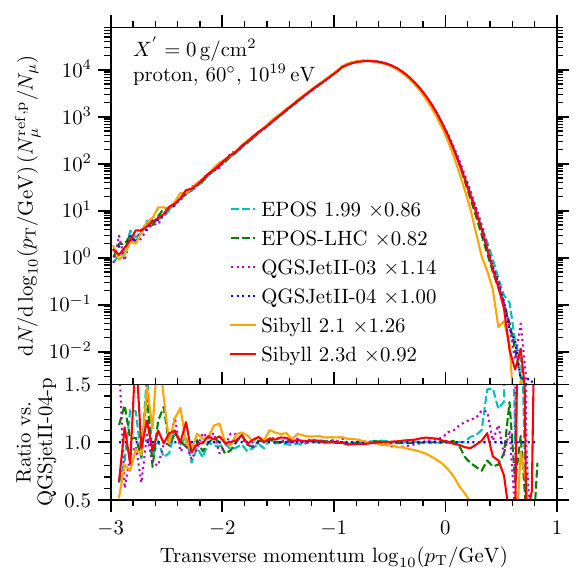}
   \qquad
   \includegraphics[width=0.45\textwidth]{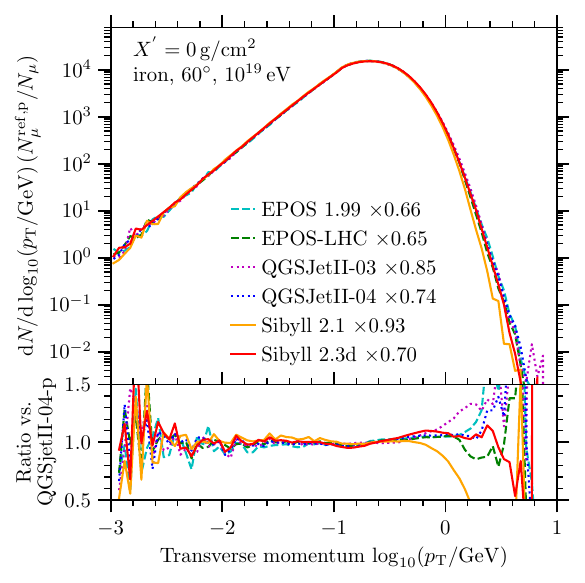}

  \caption{The average spectrum of transverse momenta of muons at the shower stage where production of muons is maximal ($X^{'}=0\,$g$/$cm$^2$). Shown are air showers simulated with different hadronic interaction models and primaries. Proton-induced showers are shown on the left, iron-induced showers are shown on the right. The spectra are re-scaled with corresponding factors shown in the labels so the total number of produced muons matches the number in proton showers simulated with \qgs. In the top of each figure the differential spectrum is shown. In the bottom the ratio to the spectrum for proton showers simulated with \qgs is shown.
  \label{fig:pt-spec-avg}}
\end{figure}

In Figure~\ref{fig:pt-spec-avg}, the average $p_{\rm T}$-spectra in showers initiated by a proton (left panel) or iron (right panel) primary are shown for different hadronic interaction models. Note that the distributions are scaled to the peak value of \qgs to better compare the shapes of the distributions and remove effects from the varying total numbers of muons between the models. The universality in the shape of the $p_{\rm T}$-spectra is even more striking than between individual showers. Note also the ratio of the spectra to \qgs that is shown at the bottom of each panel. At low-$p_{\rm T}$, the different implementations scatter randomly around the central value, indicating that these differences are merely due to undersampling or artificial fluctuations due to the thinning algorithm. At high-$p_{\rm T}$ the situation is different. Here clear trends for each model are visible. As it was argued before muon $p_{\rm T}$ comes from the last interaction which for most muons occur around the critical energy of pions. Since this is in the range of the low-energy interaction models, we do not expect the distributions to be different between high-energy models. Only when muons originate from interactions in the range of the high-energy models, the ratio shown in Figure~\ref{fig:pt-spec-avg} starts to deviate from one.

\begin{figure}
  \begin{center}

    \includegraphics[width=0.45\textwidth]{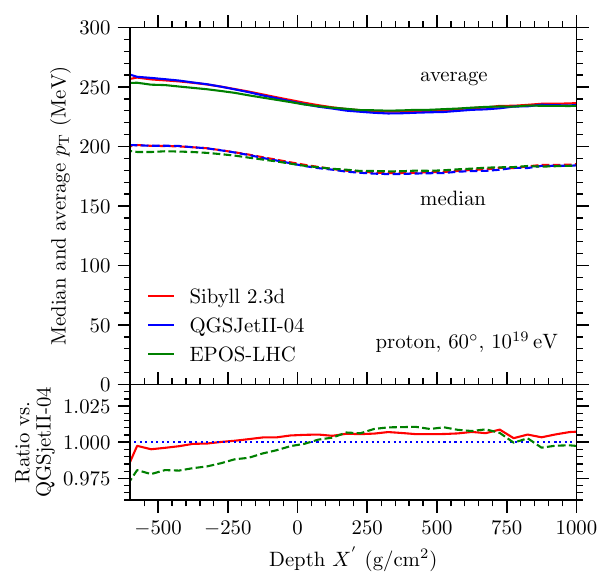}
  \caption{Evolution of average and median transverse momentum of the muons with the depth of production. Note that in $X^{'}$ the depth of maximum production is at $0\,$g$/$cm$^2$.}
  \label{fig:cpt-prof-avg-p}
 \end{center}
\end{figure}

For now, we have focused on the $p_{\rm T}$ spectra at the maximum of muon production. In Figure~\ref{fig:cpt-prof-avg-p}, the evolution of the average and median transverse momentum with atmospheric depth is shown. While both average and median decrease slightly as the shower progresses and then level-off after the maximum of production ($X^{\prime}=0\,$g$/$cm$^2$) is reached, the distance between the two stays constant, indicating that the shape of the distribution is preserved.

\begin{figure}
  \centering

  \includegraphics[width=0.95\textwidth]{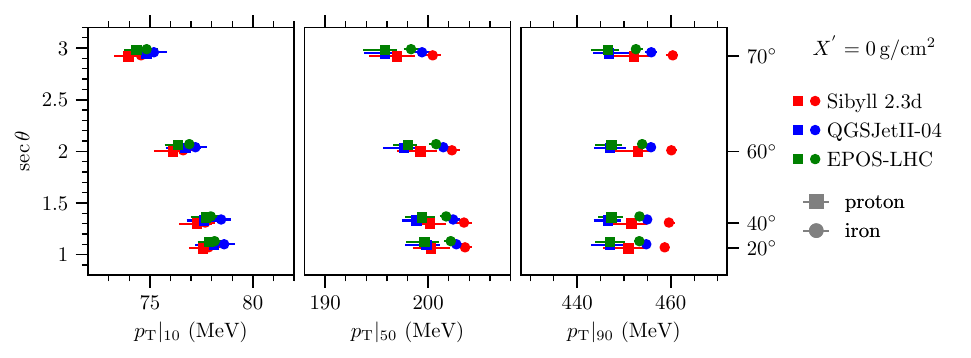}
  \caption{The quantiles of the transverse-momentum spectrum for different primary particles, zenith angles, and hadronic interaction models. The marker represents the median for the given quantile in the sample of 100 showers that were simulated. The error bars indicate the shower-to-shower fluctuations (one standard deviation) of that quantile.\label{fig:cpt-univ}}
\end{figure}

To compare the shape of the spectra quantitatively we have defined ${p_{\rm T}|_{10}}$, ${p_{\rm T}|_{50}}$, and ${p_{\rm T}|_{90}}$ as the values where the cumulative of the $p_{\rm T}$-distribution $f(p_{\rm T})=\iint f(X',E_\text{i}, p_{\rm T}) \, \mathrm{d}X^\prime \, \mathrm{d}E_\text{i}$ reaches the $q=10\%$, $q=50\%$ (median), and $q=90\%$ quantiles as
\begin{equation}
  \int_{0}^{p_{\rm T}|_q} f(p_{\rm T})\, \mathrm{d}p_{\rm T}=q \ .
\end{equation}
In Figure~\ref{fig:cpt-univ}, the evolution of the quantiles ${p_{\rm T}|_q}$ for different zenith angles, primaries, and interaction models is shown. To assess the shower-to-shower fluctuations, the quantiles were calculated for the spectra of 100 distinct showers. The symbols shown in the figure correspond to the median value of each $p_{\rm T}$ quantile over the ensemble of 100 showers, while the error bars indicate the fluctuations around that value (1$\sigma$-interval). The shower-to-shower fluctuations are larger for proton-induced showers. This is expected from the superposition model where the fluctuations for a nucleus with nucleon number $A$ are naturally suppressed by $1/\sqrt{A}$~\cite{Engel:1992vf}. In this same figure, it can be seen that the shape of the $p_{\rm T}$-distribution does not change much between the different primaries or interaction models as the change in the quantiles under these parameters is at the percent level. The only notable change is due to the zenith angle with a maximum variation of 5\% (see Appendix~\ref{appTables} for a summary of the quantiles and their variations). Meanwhile the shower-to-shower fluctuations, which in case of the median $p_{\rm T}$ are at the level of 1\% (0.3\%) for proton (iron) primaries, here vary by 10 to 30\% between different hadronic interaction models.

\begin{figure}
  \begin{center}
    \includegraphics[width=0.45\textwidth]{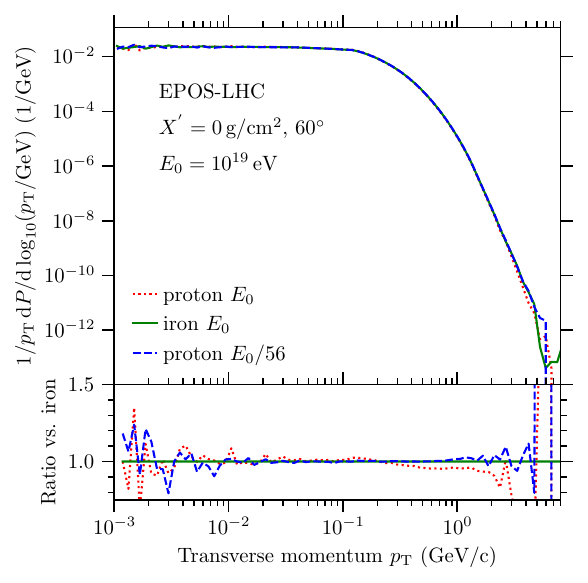}
   \caption{The average $p_{\rm T}$-spectrum for proton and iron showers. Red and green are the spectra for proton and iron showers at a primary energy of $10^{19}\,$eV. Shown in blue is the $p_{\rm T}$-spectrum of muons for a proton shower with the same primary energy per nucleon as an iron shower at $10^{19}\,$eV.}
   \label{fig:pt-spec-p-vs-fe}
\end{center}
\end{figure}

Finally, in Figure~\ref{fig:pt-spec-p-vs-fe} the change of shape of the average $p_{\rm T}$-distribution under the change of the primary particle and the primary energy is demonstrated. Also here, the shape is mostly unchanged. The only notable difference is that in lower energy showers there are slightly more muons with large transverse momenta. Through this figure, one can clearly see that the change in the shape of the distribution for different primaries is mostly due to the change in the energy per nucleon. Earlier, we argued that the $p_{\rm T}$ of muons is dominated by the last interaction. The superposition model assumes that the interaction of a primary nucleus with $A$ nucleons and energy $E_0$ is described by $A$ independent interactions of protons at the energy $E_0/A$. As seen in Figure~\ref{fig:pt-spec-p-vs-fe}, the bulks of the distributions are in agreement. 

\section{Universality of the muon production depth distributions}
\label{sec:mpd}

The Muon Production Depth distribution (MPD distribution) counts the number of muons produced in each interval of slant depth. It is equivalent to the decay rate per unit of slant depth of the mesons belonging to the hadronic cascade, and the MPD therefore corresponds to an image of the longitudinal development of the hadronic shower. In Ref.~\cite{Andringa:2011ik}, the MPD distribution was found to be well described by a Gaisser-Hillas function. The number of all muons that are produced in each slant depth interval defines the {\it total/true} MPD distributions $h(X)$. In contrast the production depth of the muons that arrive at a particular location at the ground defines the {\it apparent} MPD, which is the one accessible experimentally. Both distributions are not identical and are related through propagation effects, described in Ref.~\cite{Cazon:2012ti}. In this paper, we focus on the {\it total/true} MPD, as it is the distribution that corresponds to the hadronic cascade.

\begin{figure}
  \centering
  \includegraphics[width=0.48\textwidth]{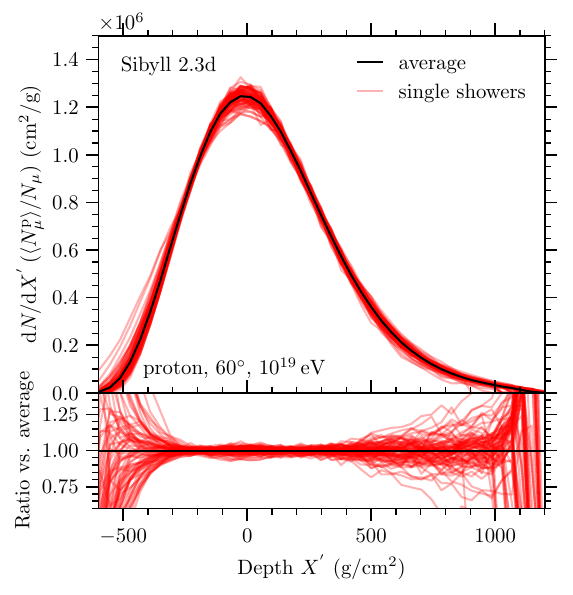}
  \hfill
    \includegraphics[width=0.48\textwidth]{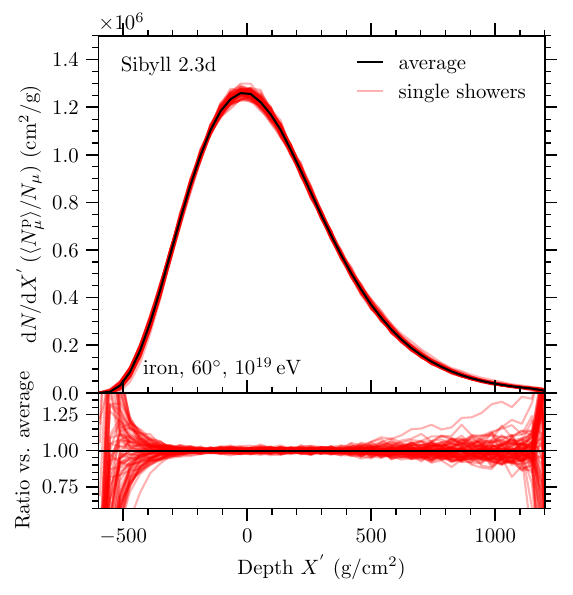}
  \caption{MPD for proton-induced (left) and iron-induced (right) air showers simulated with \sib. The primary energy is $10^{19}\,$eV, and the zenith angle is $60^{\circ}$. The profiles are shifted such that the maximum is at $0\,$g$/$cm$^2$ and scaled to the number of muons in showers induced by protons. The profiles for 100 individual showers are shown in red. The average profile of these is shown in black. \label{fig:x-prof-indiv}}
\end{figure}

To start with, in Figure~\ref{fig:x-prof-indiv} the MPD of 100 proton-initiated showers (left) and 100 iron-initiated showers (right), simulated with \sib, are shown. For each shower, $\mathcal{X}^\mu_{\rm max}$ was calculated, and all distributions were normalized to the area of the average MPD (which is ${\cal N}_\mu$). The profiles are next plotted as a function of $X'$ (shown in red). It can be seen how the shower-to-shower profile fluctuates around the average profile (shown in black). In the bottom part of the figure, the ratio of the individual shower profiles with respect to the average profile is exhibited. It can be seen that the central region around $X'=0\,$g$/$cm$^2$ maintains a high degree of universality.

\begin{figure}
  \centering

  \includegraphics[width=0.48\textwidth]{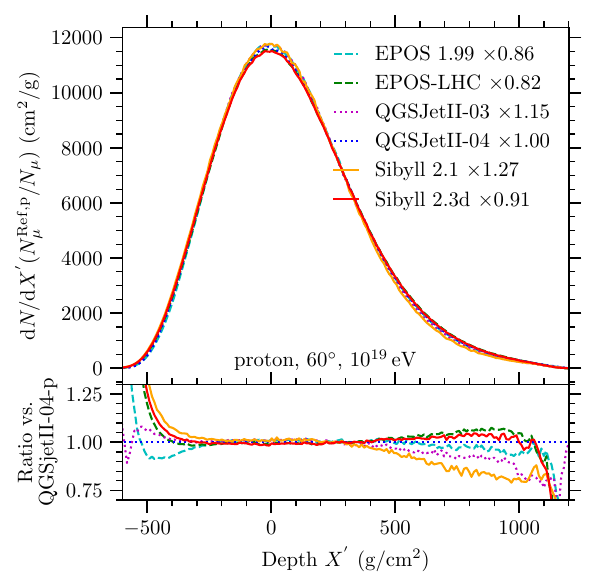}
  \hfill
  \includegraphics[width=0.48\textwidth]{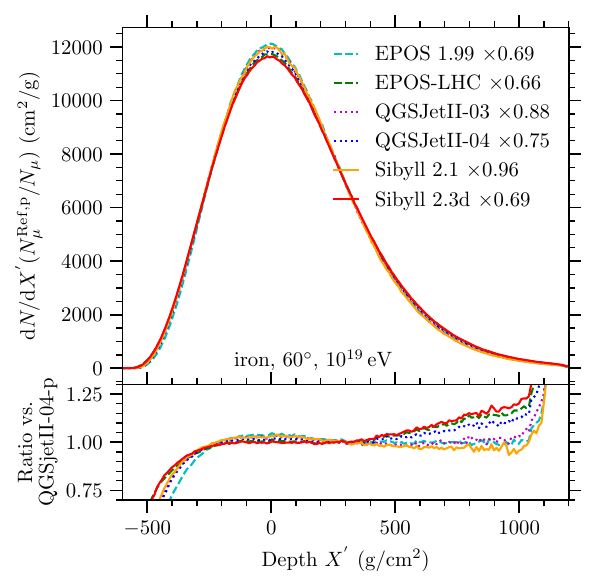}

  \caption{Average MPD for different hadronic interaction models scaled (scaling factors are given in the legend) to match the normalization of proton showers simulated with \qgs. Proton primaries are shown on the left, iron primaries are shown on the right.
    \label{fig:x-prof}
  }
  
\end{figure}

In Figure~\ref{fig:x-prof}, the average profile is plotted for different hadronic interaction models. Proton-initiated showers are shown on the left, and iron-initiated showers are on the right. To bring out the differences in the shape, the profiles are scaled to match the normalization for proton primaries simulated with \qgs (the scaling factors are shown in the legend of the figure). The bottom plot shows the ratio for proton showers simulated with \qgs. Much like the case of individual shower profiles the average profiles for different hadronic interaction models are universal in the region of the maximum of production ($X'=0\,$g$/$cm$^2$). Notable differences appear in the early and the late part of the shower development. In particular the comparison between pre- and post-LHC models shows that the overall increase of the number of muons in the new models is mainly achieved in the late shower stages. In fact, comparing proton and iron profiles we find that proton showers in the post-LHC model \qgs have the same profile shape as iron showers in the pre-LHC models. Generally, and this is similar to what is found for the EM shower profile~\cite{USP2}, muon production is more asymmetric (lower in the early stages and higher in the late stages) for iron primaries.

\begin{figure}
  \centering
  \includegraphics[width=0.95\textwidth]{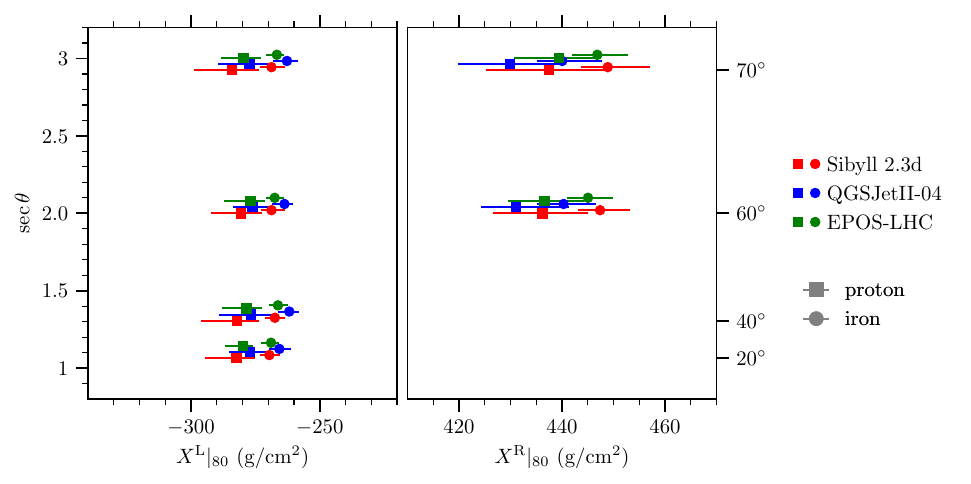}
  \caption{Zenith angle, primary mass, and model dependence of the modified quantiles of the MPD.  \label{fig:x-quant-trunc}}
\end{figure}

To make a more qualitative assessment of the degree of universality, we study the variations in the quantiles of the MPD. For showers with zenith angles below $60^\circ$ there is a good chance that, for some of the showers, the development is truncated by the ground. To remove these artificial fluctuations we only consider the MPD in the range from the start of the shower up to the shower maximum, 
and we define $X^{\rm L}|_{80}$ as the value before the maximum where the integral reaches 80\% of the total value between the start of the shower and the maximum,
\begin{equation}
  \frac
  {\int_{{X^{\rm L}}|_{80}}^0 h(X')\,\mathrm{d}X'}
 { \int_{-\infty}^0 h(X')\,\mathrm{d}X'} 
  =  0.8 \ .   \label{eq:x-quant}
\end{equation}
 For showers with zenith angles of $60^\circ$ and more, where truncation does not play a role, we also calculate $X^{\rm R}|_{80}$ which we define similar to Eq.~\eqref{eq:x-quant} but integrating from the shower maximum to infinity. Using these quantities we can study both the growth phase and the attenuating phase of the shower. In Figure~\ref{fig:x-quant-trunc}, the value of these quantiles, and their shower-to-shower fluctuations are shown for different configurations of primaries, zenith angles, and hadronic interaction models. In this figure, the median values of the quantiles are shown by the markers, and the $1\sigma$ interval of variation in the quantiles, due to the shower-to-shower fluctuations, is indicated by the error bars. Figure~\ref{fig:x-quant-trunc} confirms that within the shower-to-shower fluctuations, the shape of the MPD distribution is universal with regard to zenith angle and hadronic interaction model (total variation is at the level of 2\% or less). Between different primaries, the shape of the MPD varies at the level of 5\%. However, this variation is only marginally beyond the shower-to-shower fluctuations (1 to 4\%).

\begin{figure}
  \begin{center}
    \qquad
    \includegraphics[width=0.45\textwidth]{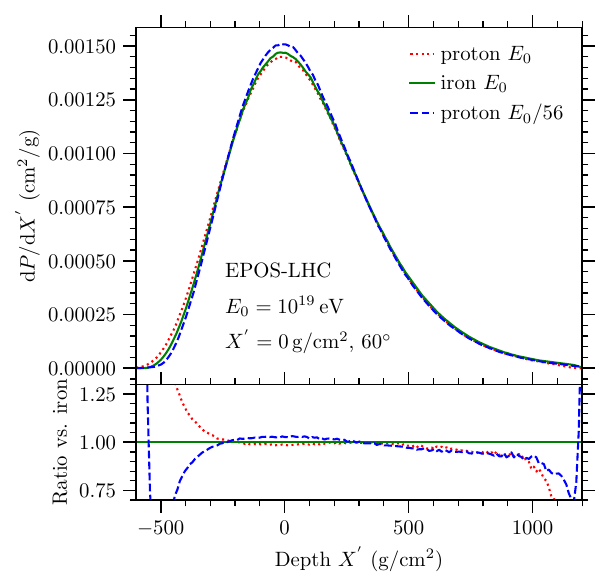}
    \caption{The average MPD for proton and iron showers. Red and green are the profiles for proton and iron showers at a primary energy of $10^{19}\,$eV. Shown in blue is the MPD for a proton shower with the same primary energy per nucleon as an iron shower at $10^{19}\,$eV.}
   \label{fig:mpd-p-vs-fe}
\end{center}
\end{figure}

The effect of the primary energy and the primary particle type on the MPD distribution is shown in Figure~\ref{fig:mpd-p-vs-fe} by comparing an average proton shower and an average iron shower at $10^{19}\,$eV with an average proton shower at $10^{17.25}\,$eV (same energy-per-nucleon as $10^{19}\,$eV iron primary). In contrast to the case of the transverse momentum distribution above, the MPD distribution is different between all three cases. In fact, proton and iron showers seem to be similar in development around the shower maximum up to $500\,$g$/$cm$^2$, while beyond that, the shape of the proton showers at different energies are similar and significantly different from iron showers. This is not surprising since iron showers are not a simple superposition of proton showers at reduced energy. In particular, the iron nucleus is expected to slowly fragment into smaller and smaller nuclei over consecutive interactions~\cite{Engel:1992vf}. This results in significant part of the hadronic cascade starting deeper in the atmosphere which explains the differences in the profiles seen above.

\section{Universality of the muon-production energy spectra}
\label{sec:en}
Finally, we examine the energy spectrum of muons at production. While the transverse momentum of muons, as discussed in Sect.~\ref{sec:pt}, is dominated by the last hadronic interaction, the energy spectrum of muons is expected to be influenced by the entire chain of interactions. The decisive property of the interactions is the energy spectrum of secondaries (or, equivalently, the distribution of longitudinal momentum). 
In addition, 

the shape of the energy spectrum will also depend on the multiplicities of the different types of particles produced. For instance, kaons, the second most numerous hadronic particle type in the shower, have a critical energy of around $850\,$GeV, while for pions, which are most numerous, the critical energy is only of the order of $100\,$GeV. This means that the kaons in the shower will influence the energy spectrum in the TeV region, as pions are very unlikely to decay and produce muons.

\begin{figure}
  \centering
  \includegraphics[width=0.45\textwidth]{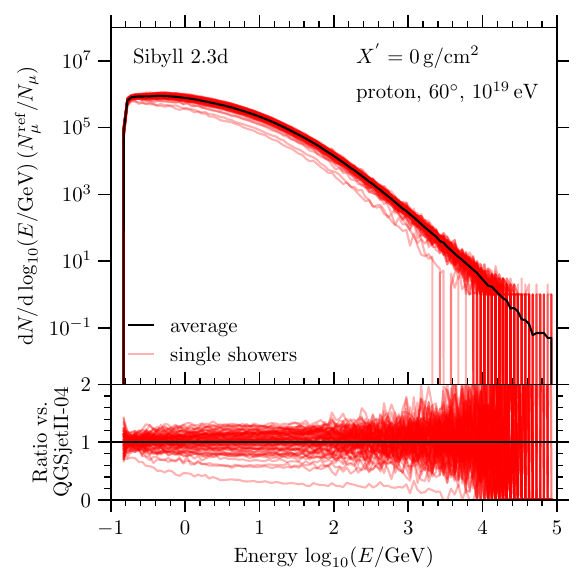}
  \qquad
  \includegraphics[width=0.45\textwidth]{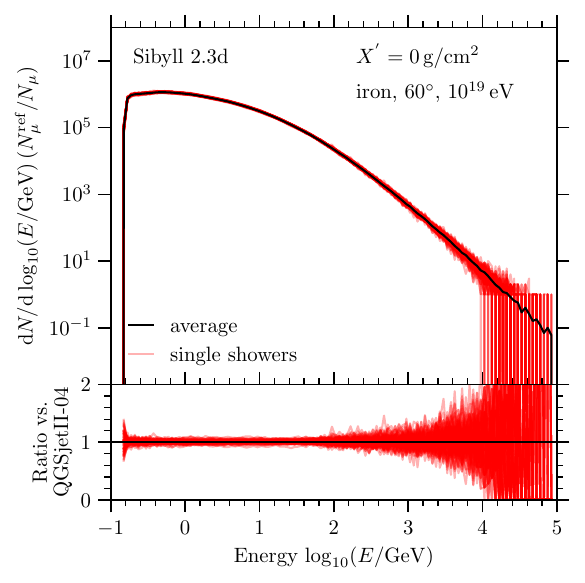}
  \caption{Energy spectrum of muons at the maximum of muon production ($X^{'}=0\,$g$/$cm$^2$) for proton-induced (left) and iron-induced (right) air showers. The primary energy is $10^{19}\,$eV and the zenith angle is $60^{\circ}$. The spectra for 100 individual showers are shown in red. The average spectrum is shown in black.
    \label{fig:en-spec-indiv}
  }
\end{figure}

In Figure~\ref{fig:en-spec-indiv}, the energy spectra for muons at the maximum ($X^{'}=0\,$g$/$cm$^2$) are shown for proton and iron-induced air showers. The average spectrum of an ensemble of 100 showers is shown in black, while the individual spectra are shown in red. The spectra here are normalized to the average number of muons of \qgs. The ratio between the individual spectra and the average spectrum in \qgs is shown in the panel at the bottom. Much like in the case of the MPD and the $p_{\rm T}$ distributions, we see that the shower-to-shower fluctuations do not significantly change the shape of the spectra.

\begin{figure}
  \centering
     \includegraphics[width=0.45\textwidth]{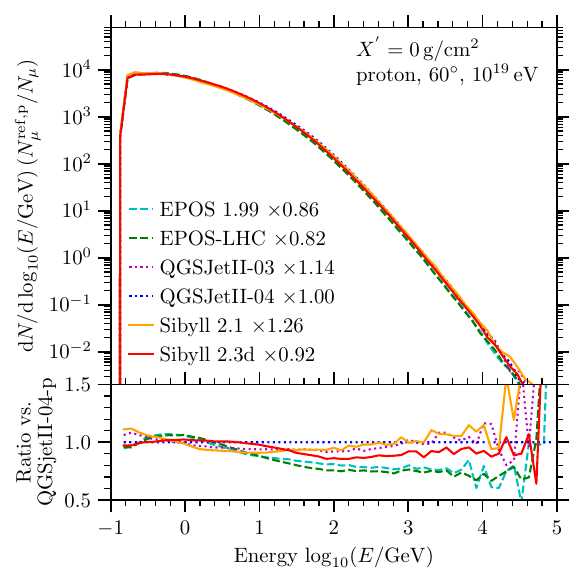}
  \qquad
     \includegraphics[width=0.45\textwidth]{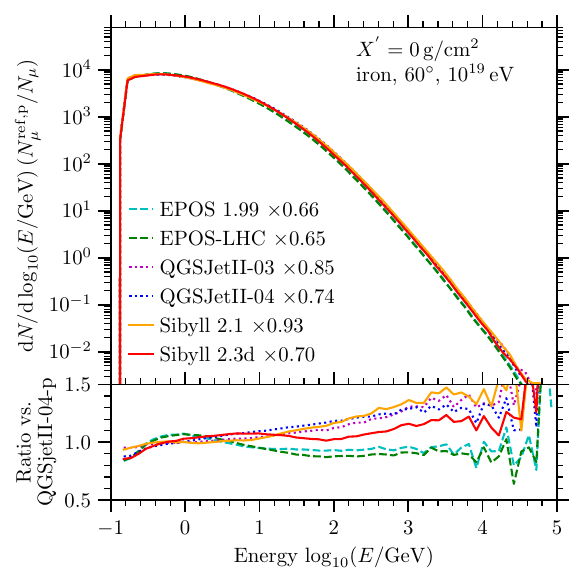}
     \caption{The average energy spectrum of muons for proton (left) and iron (right) primaries for different hadronic interaction models. Spectra are normalized to the spectrum of proton primaries in \qgs with scaling factors given in the legend. The scaling factors for each model are indicated in the legend. The bottom plot shows the ratio with \qgs.
       \label{fig:avg-en-spec}
     }
\end{figure}

The average energy spectra of muons at $X^\prime=0\,$g$/$cm$^2$ is shown in Figure~\ref{fig:avg-en-spec} for proton (left) and iron (right) primaries. These distributions are compared for pre- and post-LHC interaction models. While the overall shape of the spectra is similar between models, the ratio shown in the bottom of the plots reveals that the precise shape differs. In the low-energy region around $1\,$GeV, the differences are at the level of 10\%, increasing to 50\% or more in the high-energy region beyond a TeV. Note that, as in the case of the MPD, the relative difference between the models is similar between proton and iron primaries.

\begin{figure}
  \begin{center}

    \includegraphics[width=0.45\textwidth]{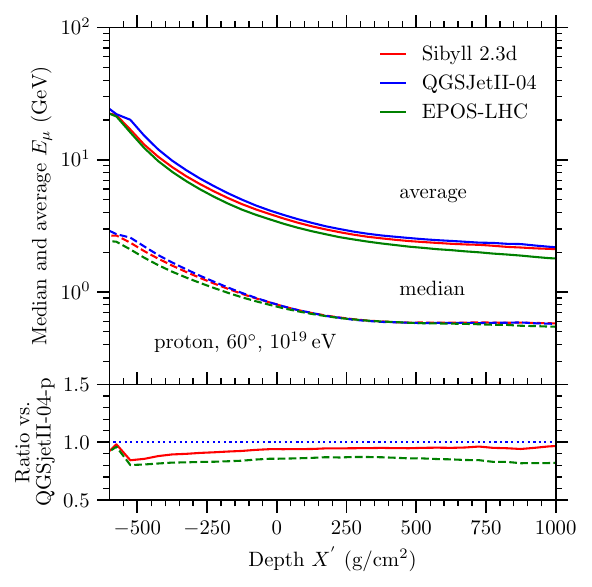}
    \qquad
    \includegraphics[width=0.45\textwidth]{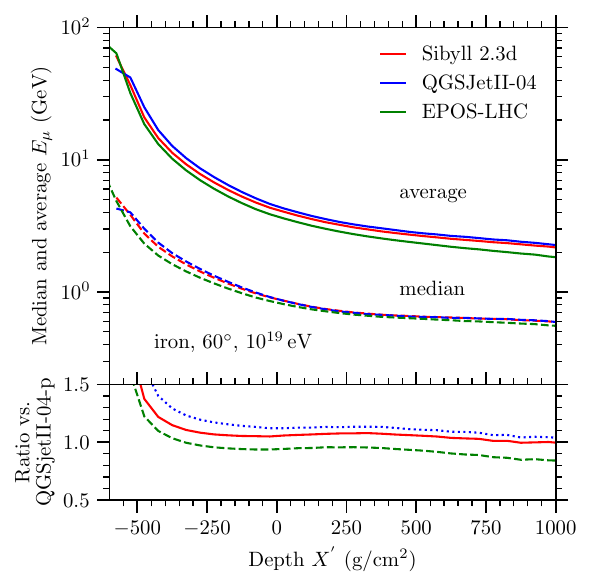}
   \caption{Evolution of average and median energy of muons with depth for proton (left) and iron (right) primaries. The bottom panel in both plots shows the ratios for the average distribution using \qgs proton as reference.}
   \label{fig:en-prof-avg}
\end{center}
\end{figure}

In Figure~\ref{fig:en-prof-avg}, the average and median energy of the muon energy spectrum are shown as a function of atmospheric depth (shifted such that the maximum of muon production is at zero). As expected, the energy of the muons continuously decreases before the shower maximum and levels off after. The analysis of the bottom plots shows that the hadronic interaction models can be distinguished by this quantity while there is no visible dependence between models with the shower depth.
Moreover, comparing proton and iron primaries, we see that the shower development seems to be universal after the maximum is reached (same median and average energy). Yet, before the shower maximum, the average and median energy of muons for iron primaries are higher.

\begin{figure}
  \centering

  \includegraphics[width=0.95\textwidth]{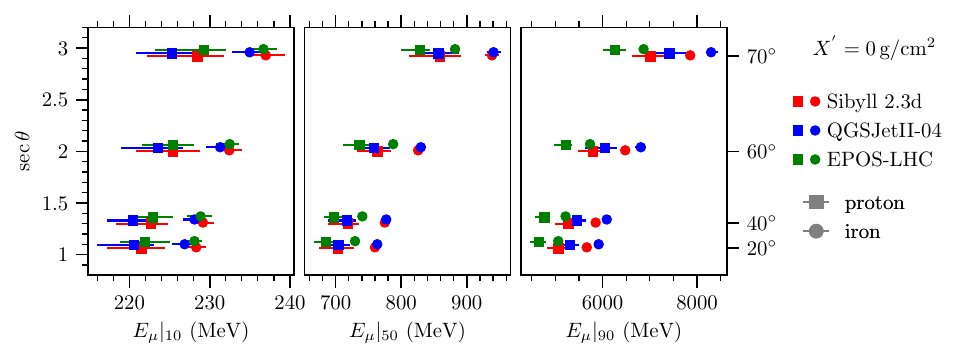}
  \caption{Primary mass, zenith angle, and model dependence of the quantiles of the energy spectrum of muons.
    \label{fig:en-quant}
  }
\end{figure}

The values of the 10\%, 50\%, and 90\% quantiles of the muon energy spectra at the shower maximum were investigated and are displayed in Figure~\ref{fig:en-quant} for different zenith angles, primaries, and the post-LHC hadronic interaction models (numerical values are in Table~\ref{tab:en} in Appendix~\ref{appTables}). The marker indicates the median value of the quantile, while the error bars show the 1$\sigma$ interval of the shower-to-shower fluctuations. As we are always looking at the quantiles at the same position in the development of the showers (at the maximum), the only thing that is changed between primaries of different zenith angles is the atmospheric density profile during the shower development. More inclined showers develop in a thinner atmosphere, and therefore the critical energy for the particles in the shower is larger. This leads to the increase of the quantiles seen in the figure (also true for the $p_{\rm T}$ quantiles in Sect.~\ref{sec:pt}). Comparing fixed zenith angles only, the variation in the shape of the energy spectrum is of the order of 6 to 10\%. Finally, comparing the energy spectra of different interaction models for a specific primary and fixed zenith angle one finds variations of the order of 3\% (5 to 7\%) for protons and iron primaries, respectively. 

\begin{figure}
  \centering
  \includegraphics[width=0.45\textwidth]{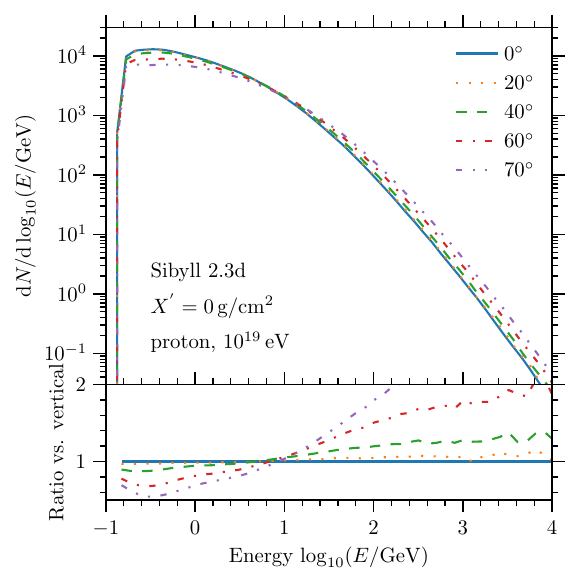}
  \qquad
  \includegraphics[width=0.45\textwidth]{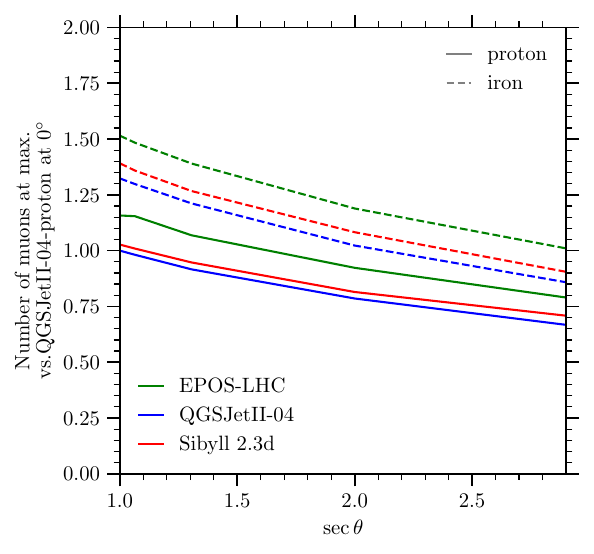}
  \caption{Variation in the muon production due to the change in the atmospheric density profile as experienced by showers of different zenith angles. On the left the energy spectrum for proton primaries is shown whereas on the right the number of muons at maximum of production is shown relative to vertical proton showers simulated with \qgs.
    \label{fig:en-spec-vs-zenith}
  }
\end{figure}

The effect of the increase of the critical energy with zenith angle is further illustrated by the hardening of the energy spectrum in Figure~\ref{fig:en-spec-vs-zenith} and the decrease in the average number of muons at the maximum relative to the number of muons in vertical showers.

\begin{figure}
  \begin{center}

    \includegraphics[width=0.45\textwidth]{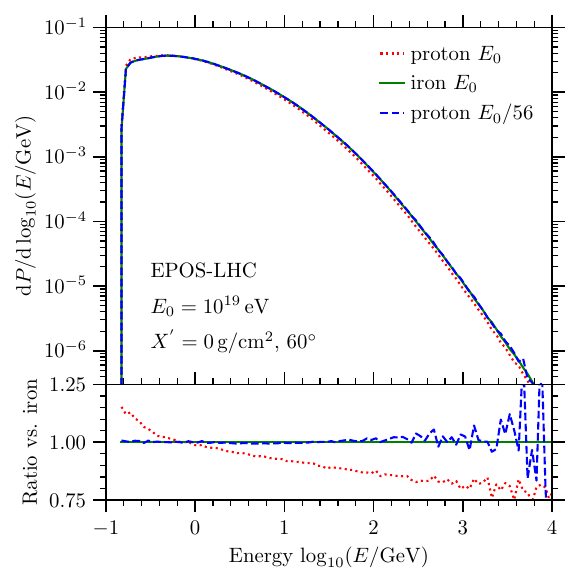}
   \caption{The average energy spectrum of muons for proton and iron showers. Red and green are the spectra for proton and iron showers at a primary energy of $10^{19}\,$eV. Shown in blue is the energy spectrum of muons for a proton shower with the same primary energy per nucleon as an iron shower at $10^{19}\,$eV}
   \label{fig:en-spec-p-vs-fe}
\end{center}
\end{figure}

Figures~\ref{fig:en-prof-avg} and~\ref{fig:en-quant} suggest that the energy spectrum for iron primaries is harder (more high-energy muons) than for proton primaries. This is confirmed in Figure~\ref{fig:en-spec-p-vs-fe} which directly compares the energy spectrum of proton and iron primaries. In addition, the energy spectrum of a proton shower with the same energy per nucleon as an iron shower at $10^{19}\,$eV, that is with a primary energy of $E_0/56~=~10^{17.25}\,$eV is shown. After accounting for this difference in initial conditions, the shapes of the energy spectra of the different primaries are identical. This is a trivial prediction of the superposition model, where the interaction of an iron nucleus is described by 56 \textit{independent} interactions of protons at a reduced energy. Note, however, that the interaction model \epos, which was used in Figure~\ref{fig:en-spec-p-vs-fe} does not use superposition but includes a more complete treatment of nuclear interactions, including nucleon-nucleon correlations and effects due to the interaction of secondary particles with the quark-gluon plasma. Since this more complex model still matches superposition closely, additional nuclear effects due to the iron primary seem to not influence the production of muons.

\section{Summary of the universality in the EAS distributions}

\begin{table}[h]
  \caption{Variation in the median of the MPD, energy spectrum, and $p_{\rm T}$ spectrum under the primary particle and the high-energy (HE) interaction model. Values are averaged over the zenith angle. The relative variation with respect to the average in percent is shown in brackets.}
  \begin{center}
    \renewcommand{\arraystretch}{1.2} \begin{tabular}{llllll}
\hline
 & \multicolumn{5}{c}{Absolute (relative) variation in median under} \\ \hline varying: & \multicolumn{2}{c}{HE model} & \multicolumn{3}{c}{primary} \\ \hline fixing: & p & Fe &EPOS-LHC&QGSJetII-04&Sibyll 2.3d \\ \hline
 $\langle \mathcal{X}^{\mu}_{\rm max} \rangle$ (g$/$cm$^2$) & $ 28$ ($4.5$) & $ 26$ ($4.7$)  & $ 73$ ($12.0$) & $ 74$ ($12.8$) & $ 82$ ($13.6$)  \\
$\langle \mathcal{N}^{*}_{\mu} \rangle$ & $1.7{\times}10^{7}$  ($16$)   & $1.9{\times}10^{7}$  ($14$)  & $2.8{\times}10^{7}$  ($21$)   & $2.6{\times}10^{7}$  ($23$)   & $2.7{\times}10^{7}$  ($23$)  \\
\arrayrulecolor{gray}\hline
$X^{\rm L}|_{80}$ (g$/$cm$^2$) & $  5$ ($1.8$) & $  5$ ($1.9$)  & $ 11$ ($4.1$) & $ 13$ ($4.9$) & $ 13$ ($4.8$)  \\
$E_{\mu}|_{50}$ (MeV) & $ 29$ ($3.6$) & $ 51$ ($5.8$)  & $ 54$ ($6.7$) & $ 77$ ($9.1$) & $ 72$ ($8.5$)  \\
$p_{\rm T}|_{50}$ (MeV) & $  1$ ($0.6$) & $  2$ ($0.9$)  & $  3$ ($1.3$) & $  4$ ($1.8$) & $  3$ ($1.6$)  \\
\arrayrulecolor{black}\hline 
\end{tabular}

  \end{center}
  \label{tab:summary}
\end{table}
In Table~\ref{tab:summary}, we summarize the universality in the shape and the normalisation of the distributions by showing the variation in the median between different primaries and high-energy (HE) interaction models. We fix one of these two parameters (primary or HE model) and vary the other while averaging over the zenith angle. Both the maximum variation within the ensemble and the variation relative to the average are shown. The largest deviation from universality are observed in the energy spectrum of muons which are at the level of 4\% (proton) to 6\% (iron) between interaction models and 7\% (\epos) to 9\% (\qgs) between primaries. For the $p_{\rm T}$-spectrum universality holds within 2\% and the MPD within 5\%. A complete listing of the variations including the variation in the shower-to-shower fluctuations is given in Tables~\ref{tab:en}--\ref{tab:delta-sh2sh-x} in Appendix~\ref{appTables}.

\section{Final remarks and summary}
The production of muons in extensive air showers can be described by two quantities, which are the total number of muons $\mathcal{N}_\mu$ and the depth $\mathcal{X}_{\rm max}^\mu$ at which the maximum of production is reached, and by a distribution with three variables, the transverse momentum $p_{\rm T}$, the depth of production $X'$ and the energy $E_\mu$ at production. Measurements of the maximum of the {\it apparent} muon production depth ({\it apparent} MPD), $X_{\rm max}^\mu$ are at variance with the predictions from the current hadronic interaction models~\cite{Aab:2014dua}, which have been shown to be very sensitive to the diffraction cross section of pions and the baryon content of the shower~\cite{Pierog:2006qv,sergey}. 
On the other hand, the muon number has been shown to be at a deficit in simulations with all hadronic models with respect to the measured data -- \emph{the EAS muon puzzle}. Various attempts have been made to explain this deficit, involving the baryon content, enhanced $\rho^0$ or strangeness production~\cite{Grieder:1973x1,Pierog:2006qv,Ostapchenko:2013pia,Drescher:2007hc} or more exotic scenarios~\cite{Aloisio:2014dua,AlvarezMuniz:2012dd,Anchordoqui:2016oxy,Farrar:2013sfa}.
In this paper, we discussed the universality of the distribution of the transverse momentum, the depth of production, and the energy at production of the muons when referred to the depth where the muon production reaches its maximum.  We assess the degree of universality by studying how much the shape of these three distributions varies with zenith angle, primary particle, primary energy, and hadronic interaction model. We have shown that the transverse-momentum and production-depth distributions are fairly universal with a maximal variation of $4\,$MeV$/c$ (2\% relative variation) in case of the $p_{\rm T}$ spectrum and $13\,$g$/$cm$^2$ (5\% relative variation) in case of the MPD. The most significant deviations from universality are seen in the energy spectrum between primaries which varies by $50\,$MeV to $80\,$MeV (7\% to 9\%). It is worth noting that the reported universalities are present in the simulations for the family of models considered in this work. They should be evaluated using cosmic-ray data,  which hopefully will be done in the coming years thanks to the detector upgrades that allow to separately measure the shower electromagnetic and muonic components at the ground~\cite{AugerPrime}, and relate the latter to the muon distributions at production.

The importance of the loss of universality in the number of muons is twofold: since the response of many detectors depends on the energy of the muons, the comparison of the recorded signal to simulations with a different muon spectrum might result in a wrong interpretation. On the other hand, and from the fundamental point of view, we do not know yet what is the hadronic interaction physics to correctly describe the muon number or the depth of maximum production. A combination of the effects listed above may be sufficient but it is not clear which. The exact impact of the particular extension on the energy spectrum of muons will have to be done in the future. Here, we argue that the shower displays a universal behaviour of most quantities related to the muon component and that the only significant differences across models arise in the muon energy spectrum, particularly in the high-energy tails. Turning the argument around, a measurement of the muon spectrum might give a very strong indication as to the solution of the muon puzzle.

\acknowledgments

The authors would like to thank the colleagues from the Pierre~Auger~Collaboration for all the fruitful discussions. The authors would like to thank Sofia Andringa, Armando di Matteo, Marvin Gottowik, Gonzalo Parente, Miguel Martins, Darko Veberič and M\'ario Pimenta for the comments and suggestions on the manuscript. 
L.~C. acknowledges the financial support from Xunta de Galicia (Centro singular de investigación de Galicia accreditation 2019-2022), grant ED431F 2022/15 and ED431F 2022/15, by European Union ERDF, and by the “María de Maeztu” Units of Excellence program MDM-2016-0692 and the Spanish Research State Agency, grant PID2019-105544GB-I00 and program "Ramon y Cajal", Grant No. RYC2019-027017-I.
This work has also been financed by national funds through FCT - Fundação para a Ciência e a Tecnologia, I.P., under project CERN/FIS-PAR/0020/2021. R.~C.\ is grateful for the financial support by OE - Portugal, FCT, I. P., under DL57/2016/cP1330/cT0002. F.~R.\ has received funding from the European Union’s Horizon 2020 research and innovation programme under the Marie Skłodowska-Curie grant agreement No 101065027.

\bibliographystyle{JHEP}


\providecommand{\href}[2]{#2}\begingroup\raggedright\endgroup


\appendix

\section{Quantiles of the distributions at production and their variation}
\label{appTables}
\begin{table}
  \caption{10\%, 50\%, 90\% quantiles and mean of the muon energy spectrum for different cosmic rays. Values are median for 100 showers with 1$\sigma$ interval of the shower-to-shower fluctuations.}
  \begin{center}
    \begin{tabular}{lllllll}
\hline
 \multicolumn{7}{c}{Energy spectrum (GeV)} \\ \hline 
 $\theta$ & model & primary & 10\% & 50\% & 90\% & mean \\ 
\hline
\multirow{6}{*}{$0^\circ$} & \multirow{2}{*}{EPOS-LHC} & p & $ 0.222 \pm 0.003 $
& $ 0.69 \pm 0.02 $
& $ 4.6 \pm 0.2 $
& $ 0.86 \pm 0.02 $
\\
&  & Fe & $ 0.228 \pm 0.001 $
& $ 0.73 \pm 0.01 $
& $ 5.1 \pm 0.1 $
& $ 0.92 \pm 0.01 $
\\
& \multirow{2}{*}{QGSJetII-04} & p & $ 0.221 \pm 0.004 $
& $ 0.70 \pm 0.02 $
& $ 5.3 \pm 0.2 $
& $ 0.91 \pm 0.03 $
\\
&  & Fe & $ 0.227 \pm 0.002 $
& $ 0.76 \pm 0.01 $
& $ 5.9 \pm 0.1 $
& $ 0.98 \pm 0.01 $
\\
& \multirow{2}{*}{Sibyll 2.3d} & p & $ 0.221 \pm 0.004 $
& $ 0.70 \pm 0.03 $
& $ 5.1 \pm 0.2 $
& $ 0.90 \pm 0.03 $
\\
&  & Fe & $ 0.228 \pm 0.001 $
& $ 0.76 \pm 0.01 $
& $ 5.7 \pm 0.1 $
& $ 0.96 \pm 0.01 $
\\
\\
\multirow{6}{*}{$20^\circ$} & \multirow{2}{*}{EPOS-LHC} & p & $ 0.223 \pm 0.002 $
& $ 0.70 \pm 0.01 $
& $ 4.8 \pm 0.2 $
& $ 0.88 \pm 0.02 $
\\
&  & Fe & $ 0.229 \pm 0.002 $
& $ 0.74 \pm 0.01 $
& $ 5.2 \pm 0.1 $
& $ 0.93 \pm 0.01 $
\\
& \multirow{2}{*}{QGSJetII-04} & p & $ 0.220 \pm 0.003 $
& $ 0.72 \pm 0.02 $
& $ 5.5 \pm 0.2 $
& $ 0.92 \pm 0.03 $
\\
&  & Fe & $ 0.228 \pm 0.001 $
& $ 0.78 \pm 0.01 $
& $ 6.1 \pm 0.1 $
& $ 0.99 \pm 0.01 $
\\
& \multirow{2}{*}{Sibyll 2.3d} & p & $ 0.223 \pm 0.003 $
& $ 0.72 \pm 0.02 $
& $ 5.3 \pm 0.2 $
& $ 0.92 \pm 0.03 $
\\
&  & Fe & $ 0.229 \pm 0.001 $
& $ 0.77 \pm 0.01 $
& $ 5.9 \pm 0.1 $
& $ 0.98 \pm 0.01 $
\\
\\
\multirow{6}{*}{$40^\circ$} & \multirow{2}{*}{EPOS-LHC} & p & $ 0.225 \pm 0.003 $
& $ 0.74 \pm 0.02 $
& $ 5.2 \pm 0.2 $
& $ 0.93 \pm 0.02 $
\\
&  & Fe & $ 0.232 \pm 0.001 $
& $ 0.79 \pm 0.01 $
& $ 5.7 \pm 0.1 $
& $ 0.99 \pm 0.01 $
\\
& \multirow{2}{*}{QGSJetII-04} & p & $ 0.224 \pm 0.004 $
& $ 0.76 \pm 0.02 $
& $ 6.0 \pm 0.3 $
& $ 0.97 \pm 0.03 $
\\
&  & Fe & $ 0.231 \pm 0.001 $
& $ 0.83 \pm 0.01 $
& $ 6.8 \pm 0.1 $
& $ 1.06 \pm 0.01 $
\\
& \multirow{2}{*}{Sibyll 2.3d} & p & $ 0.225 \pm 0.004 $
& $ 0.76 \pm 0.03 $
& $ 5.8 \pm 0.3 $
& $ 0.97 \pm 0.03 $
\\
&  & Fe & $ 0.232 \pm 0.002 $
& $ 0.83 \pm 0.01 $
& $ 6.5 \pm 0.1 $
& $ 1.04 \pm 0.01 $
\\
\\
\multirow{6}{*}{$60^\circ$} & \multirow{2}{*}{EPOS-LHC} & p & $ 0.229 \pm 0.004 $
& $ 0.83 \pm 0.02 $
& $ 6.3 \pm 0.2 $
& $ 1.03 \pm 0.03 $
\\
&  & Fe & $ 0.237 \pm 0.002 $
& $ 0.88 \pm 0.01 $
& $ 6.9 \pm 0.1 $
& $ 1.10 \pm 0.01 $
\\
& \multirow{2}{*}{QGSJetII-04} & p & $ 0.225 \pm 0.004 $
& $ 0.86 \pm 0.03 $
& $ 7.4 \pm 0.4 $
& $ 1.09 \pm 0.04 $
\\
&  & Fe & $ 0.235 \pm 0.002 $
& $ 0.94 \pm 0.01 $
& $ 8.3 \pm 0.1 $
& $ 1.19 \pm 0.01 $
\\
& \multirow{2}{*}{Sibyll 2.3d} & p & $ 0.228 \pm 0.005 $
& $ 0.86 \pm 0.04 $
& $ 7.0 \pm 0.4 $
& $ 1.08 \pm 0.04 $
\\
&  & Fe & $ 0.237 \pm 0.002 $
& $ 0.94 \pm 0.01 $
& $ 7.9 \pm 0.1 $
& $ 1.17 \pm 0.01 $
\\
\\
\multirow{6}{*}{$70^\circ$} & \multirow{2}{*}{EPOS-LHC} & p & $ 0.229 \pm 0.004 $
& $ 0.90 \pm 0.03 $
& $ 7.4 \pm 0.4 $
& $ 1.12 \pm 0.04 $
\\
&  & Fe & $ 0.238 \pm 0.003 $
& $ 0.98 \pm 0.01 $
& $ 8.3 \pm 0.1 $
& $ 1.21 \pm 0.01 $
\\
& \multirow{2}{*}{QGSJetII-04} & p & $ 0.220 \pm 0.010 $
& $ 0.95 \pm 0.05 $
& $ 9.0 \pm 1.0 $
& $ 1.2 \pm 0.10 $
\\
&  & Fe & $ 0.236 \pm 0.003 $
& $ 1.06 \pm 0.02 $
& $ 10.3 \pm 0.2 $
& $ 1.33 \pm 0.02 $
\\
& \multirow{2}{*}{Sibyll 2.3d} & p & $ 0.230 \pm 0.010 $
& $ 0.90 \pm 0.10 $
& $ 8.4 \pm 0.5 $
& $ 1.20 \pm 0.10 $
\\
&  & Fe & $ 0.238 \pm 0.002 $
& $ 1.05 \pm 0.01 $
& $ 9.6 \pm 0.2 $
& $ 1.31 \pm 0.02 $
\\
\\
\hline
\end{tabular}

  \end{center}
  \label{tab:en}
\end{table}

\begin{table}
  \caption{10\%, 50\%, 90\% quantiles and mean of the transverse momentum spectrum for different cosmic rays. Values are median for 100 showers with 1$\sigma$ interval of the shower-to-shower fluctuations. For the median shower-to-shower fluctuations are at the level of $1\,$\% or less.}
  \begin{center}
    \begin{tabular}{lllllll}
\hline
 \multicolumn{7}{c}{Transverse momentum spectrum (GeV)} \\ \hline 
 $\theta$ & model & primary & 10\% & 50\% & 90\% & mean \\ 
\hline
\multirow{6}{*}{$0^\circ$} & \multirow{2}{*}{EPOS-LHC} & p & $ 0.078 \pm 0.001 $
& $ 0.200 \pm 0.002 $
& $ 0.447 \pm 0.003 $
& $ 0.190 \pm 0.001 $
\\
&  & Fe & $ 0.078 \pm 0.001 $
& $ 0.202 \pm 0.001 $
& $ 0.453 \pm 0.001 $
& $ 0.192 \pm 0.001 $
\\
& \multirow{2}{*}{QGSJetII-04} & p & $ 0.078 \pm 0.001 $
& $ 0.200 \pm 0.002 $
& $ 0.447 \pm 0.004 $
& $ 0.191 \pm 0.001 $
\\
&  & Fe & $ 0.079 \pm 0.001 $
& $ 0.203 \pm 0.001 $
& $ 0.455 \pm 0.001 $
& $ 0.193 \pm 0.001 $
\\
& \multirow{2}{*}{Sibyll 2.3d} & p & $ 0.078 \pm 0.001 $
& $ 0.200 \pm 0.002 $
& $ 0.451 \pm 0.005 $
& $ 0.191 \pm 0.001 $
\\
&  & Fe & $ 0.078 \pm 0.001 $
& $ 0.204 \pm 0.001 $
& $ 0.459 \pm 0.001 $
& $ 0.193 \pm 0.001 $
\\
\\
\multirow{6}{*}{$20^\circ$} & \multirow{2}{*}{EPOS-LHC} & p & $ 0.078 \pm 0.001 $
& $ 0.199 \pm 0.001 $
& $ 0.447 \pm 0.003 $
& $ 0.190 \pm 0.001 $
\\
&  & Fe & $ 0.078 \pm 0.001 $
& $ 0.202 \pm 0.001 $
& $ 0.453 \pm 0.001 $
& $ 0.192 \pm 0.001 $
\\
& \multirow{2}{*}{QGSJetII-04} & p & $ 0.078 \pm 0.001 $
& $ 0.199 \pm 0.001 $
& $ 0.447 \pm 0.003 $
& $ 0.190 \pm 0.001 $
\\
&  & Fe & $ 0.078 \pm 0.001 $
& $ 0.202 \pm 0.001 $
& $ 0.455 \pm 0.001 $
& $ 0.193 \pm 0.001 $
\\
& \multirow{2}{*}{Sibyll 2.3d} & p & $ 0.077 \pm 0.001 $
& $ 0.200 \pm 0.002 $
& $ 0.452 \pm 0.004 $
& $ 0.191 \pm 0.001 $
\\
&  & Fe & $ 0.078 \pm 0.001 $
& $ 0.203 \pm 0.001 $
& $ 0.460 \pm 0.001 $
& $ 0.193 \pm 0.001 $
\\
\\
\multirow{6}{*}{$40^\circ$} & \multirow{2}{*}{EPOS-LHC} & p & $ 0.076 \pm 0.001 $
& $ 0.198 \pm 0.001 $
& $ 0.447 \pm 0.003 $
& $ 0.189 \pm 0.001 $
\\
&  & Fe & $ 0.077 \pm 0.001 $
& $ 0.201 \pm 0.001 $
& $ 0.454 \pm 0.001 $
& $ 0.191 \pm 0.001 $
\\
& \multirow{2}{*}{QGSJetII-04} & p & $ 0.077 \pm 0.001 $
& $ 0.198 \pm 0.002 $
& $ 0.447 \pm 0.003 $
& $ 0.189 \pm 0.001 $
\\
&  & Fe & $ 0.077 \pm 0.001 $
& $ 0.201 \pm 0.001 $
& $ 0.456 \pm 0.001 $
& $ 0.192 \pm 0.001 $
\\
& \multirow{2}{*}{Sibyll 2.3d} & p & $ 0.076 \pm 0.001 $
& $ 0.199 \pm 0.002 $
& $ 0.453 \pm 0.004 $
& $ 0.190 \pm 0.002 $
\\
&  & Fe & $ 0.077 \pm 0.001 $
& $ 0.202 \pm 0.001 $
& $ 0.460 \pm 0.001 $
& $ 0.192 \pm 0.001 $
\\
\\
\multirow{6}{*}{$60^\circ$} & \multirow{2}{*}{EPOS-LHC} & p & $ 0.074 \pm 0.001 $
& $ 0.196 \pm 0.002 $
& $ 0.447 \pm 0.003 $
& $ 0.187 \pm 0.001 $
\\
&  & Fe & $ 0.075 \pm 0.001 $
& $ 0.198 \pm 0.001 $
& $ 0.453 \pm 0.001 $
& $ 0.189 \pm 0.001 $
\\
& \multirow{2}{*}{QGSJetII-04} & p & $ 0.075 \pm 0.001 $
& $ 0.196 \pm 0.002 $
& $ 0.447 \pm 0.004 $
& $ 0.187 \pm 0.001 $
\\
&  & Fe & $ 0.075 \pm 0.001 $
& $ 0.199 \pm 0.001 $
& $ 0.456 \pm 0.001 $
& $ 0.190 \pm 0.001 $
\\
& \multirow{2}{*}{Sibyll 2.3d} & p & $ 0.074 \pm 0.001 $
& $ 0.197 \pm 0.002 $
& $ 0.452 \pm 0.004 $
& $ 0.188 \pm 0.002 $
\\
&  & Fe & $ 0.075 \pm 0.001 $
& $ 0.200 \pm 0.001 $
& $ 0.460 \pm 0.001 $
& $ 0.190 \pm 0.001 $
\\
\\
\multirow{6}{*}{$70^\circ$} & \multirow{2}{*}{EPOS-LHC} & p & $ 0.073 \pm 0.001 $
& $ 0.194 \pm 0.002 $
& $ 0.444 \pm 0.004 $
& $ 0.185 \pm 0.001 $
\\
&  & Fe & $ 0.073 \pm 0.001 $
& $ 0.197 \pm 0.001 $
& $ 0.451 \pm 0.002 $
& $ 0.187 \pm 0.001 $
\\
& \multirow{2}{*}{QGSJetII-04} & p & $ 0.073 \pm 0.001 $
& $ 0.194 \pm 0.002 $
& $ 0.446 \pm 0.004 $
& $ 0.185 \pm 0.002 $
\\
&  & Fe & $ 0.074 \pm 0.001 $
& $ 0.198 \pm 0.001 $
& $ 0.455 \pm 0.002 $
& $ 0.188 \pm 0.001 $
\\
& \multirow{2}{*}{Sibyll 2.3d} & p & $ 0.073 \pm 0.001 $
& $ 0.195 \pm 0.002 $
& $ 0.45 \pm 0.01 $
& $ 0.186 \pm 0.002 $
\\
&  & Fe & $ 0.073 \pm 0.001 $
& $ 0.199 \pm 0.001 $
& $ 0.459 \pm 0.002 $
& $ 0.188 \pm 0.001 $
\\
\\
\hline
\end{tabular}

  \end{center}
  \label{tab:pt}
\end{table}

\begin{table}
  \caption{Pseudo-quantiles of the muon production depth for different cosmic rays. Since the shower development is cut short for inclinations below $60^\circ$ by the ground (here at $1400\,$m above sea-level) the quantiles are calculated from the maximum to the start of the shower. Values are median for 100 showers with 1$\sigma$ interval of the shower-to-shower fluctuations.}
  \begin{center}
    \begin{tabular}{lllll}
\hline
 \multicolumn{5}{c}{ muon production depth (g$/$cm$^2$) } \\ \hline 
 $\theta$ & model & primary & $X^{\rm L}|_{80}$ & $X^{\rm R}|_{80}$ \\ 
\hline
\multirow{6}{*}{$0^\circ$} & \multirow{2}{*}{EPOS-LHC} & p & $ -280 \pm 8 $
& $ 175 \pm 39 $
\\
&  & Fe & $ -269 \pm 4 $
& $ 224 \pm 11 $
\\
& \multirow{2}{*}{QGSJetII-04} & p & $ -280 \pm 10 $
& $ 185 \pm 41 $
\\
&  & Fe & $ -267 \pm 5 $
& $ 238 \pm 15 $
\\
& \multirow{2}{*}{Sibyll 2.3d} & p & $ -283 \pm 10 $
& $ 169 \pm 46 $
\\
&  & Fe & $ -271 \pm 5 $
& $ 229 \pm 13 $
\\
\\
\multirow{6}{*}{$20^\circ$} & \multirow{2}{*}{EPOS-LHC} & p & $ -280 \pm 6 $
& $ 214 \pm 29 $
\\
&  & Fe & $ -269 \pm 4 $
& $ 261 \pm 11 $
\\
& \multirow{2}{*}{QGSJetII-04} & p & $ -277 \pm 8 $
& $ 233 \pm 36 $
\\
&  & Fe & $ -266 \pm 4 $
& $ 274 \pm 13 $
\\
& \multirow{2}{*}{Sibyll 2.3d} & p & $ -282 \pm 10 $
& $ 218 \pm 38 $
\\
&  & Fe & $ -270 \pm 4 $
& $ 265 \pm 11 $
\\
\\
\multirow{6}{*}{$40^\circ$} & \multirow{2}{*}{EPOS-LHC} & p & $ -278 \pm 8 $
& $ 335 \pm 25 $
\\
&  & Fe & $ -266 \pm 4 $
& $ 366 \pm 6 $
\\
& \multirow{2}{*}{QGSJetII-04} & p & $ -277 \pm 10 $
& $ 344 \pm 27 $
\\
&  & Fe & $ -262 \pm 4 $
& $ 372 \pm 6 $
\\
& \multirow{2}{*}{Sibyll 2.3d} & p & $ -282 \pm 11 $
& $ 333 \pm 28 $
\\
&  & Fe & $ -267 \pm 4 $
& $ 369 \pm 7 $
\\
\\
\multirow{6}{*}{$60^\circ$} & \multirow{2}{*}{EPOS-LHC} & p & $ -277 \pm 8 $
& $ 437 \pm 7 $
\\
&  & Fe & $ -268 \pm 3 $
& $ 445 \pm 5 $
\\
& \multirow{2}{*}{QGSJetII-04} & p & $ -276 \pm 8 $
& $ 431 \pm 9 $
\\
&  & Fe & $ -264 \pm 4 $
& $ 440 \pm 6 $
\\
& \multirow{2}{*}{Sibyll 2.3d} & p & $ -281 \pm 10 $
& $ 436 \pm 9 $
\\
&  & Fe & $ -269 \pm 5 $
& $ 447 \pm 5 $
\\
\\
\multirow{6}{*}{$70^\circ$} & \multirow{2}{*}{EPOS-LHC} & p & $ -280 \pm 8 $
& $ 439 \pm 8 $
\\
&  & Fe & $ -267 \pm 4 $
& $ 447 \pm 5 $
\\
& \multirow{2}{*}{QGSJetII-04} & p & $ -277 \pm 10 $
& $ 430 \pm 10 $
\\
&  & Fe & $ -263 \pm 5 $
& $ 440 \pm 6 $
\\
& \multirow{2}{*}{Sibyll 2.3d} & p & $ -284 \pm 13 $
& $ 437 \pm 12 $
\\
&  & Fe & $ -269 \pm 5 $
& $ 449 \pm 7 $
\\
\\
\hline
\end{tabular}

  \end{center}
  \label{tab:x}
\end{table}


\begin{table}
  \caption{Variation in the median of the muon energy spectrum. The variation is calculated as the difference between the maximum and minimum values in a set over the average in that set. The largest variation is due to the change in the critical energy between different zenith angles. The smallest level is reached for proton primaries when varying the hadronic interaction model.}
  \begin{center}
    \begin{tabular}{llllll}
\hline
 \multicolumn{5}{c}{ Variation in $E_{\mu}|_{50}$ ($\delta_\text{total}$= 45.8 \%)} \\ \hline 
 $\theta$ & $\delta_{\text{primary}, \, \text{model}}$ (\%) & primary & $\delta_\text{model}$ (\%) & model & $\delta_\text{primary}$ (\%) \\ 
\hline
\multirow{5}{*}{$0^\circ$ } & \multirow{5}{*}{10.79 } &   p &  2.7 & & \\
 & &   Fe &  4.5 & & \\
 & & & & EPOS-LHC &  6.3 \\
 & & & & QGSJetII-04 &  8.1 \\
 & & & & Sibyll 2.3d &  7.7 \\
\hline 
\multirow{5}{*}{$20^\circ$ } & \multirow{5}{*}{10.77 } &   p &  3.0 & & \\
 & &   Fe &  4.8 & & \\
 & & & & EPOS-LHC &  6.0 \\
 & & & & QGSJetII-04 &  8.0 \\
 & & & & Sibyll 2.3d &  7.5 \\
\hline 
\multirow{5}{*}{$40^\circ$ } & \multirow{5}{*}{11.93 } &   p &  3.6 & & \\
 & &   Fe &  5.2 & & \\
 & & & & EPOS-LHC &  6.7 \\
 & & & & QGSJetII-04 &  9.0 \\
 & & & & Sibyll 2.3d &  7.8 \\
\hline 
\multirow{5}{*}{$60^\circ$ } & \multirow{5}{*}{12.70 } &   p &  3.6 & & \\
 & &   Fe &  6.4 & & \\
 & & & & EPOS-LHC &  6.3 \\
 & & & & QGSJetII-04 &  9.4 \\
 & & & & Sibyll 2.3d &  8.9 \\
\hline 
\multirow{5}{*}{$70^\circ$ } & \multirow{5}{*}{16.11 } &   p &  4.9 & & \\
 & &   Fe &  8.0 & & \\
 & & & & EPOS-LHC &  8.0 \\
 & & & & QGSJetII-04 & 11.2 \\
 & & & & Sibyll 2.3d & 10.5 \\
\hline 
\end{tabular}

  \end{center}
  \label{tab:delta-en}
\end{table}

\begin{table}
  \caption{Variation in the median of the transverse-momentum spectrum. The largest variation in the $p_{\rm T}$-spectrum is due to the change in critical energy. A variation at the level of $1\,$\% is seen with primary mass. The variation between hadronic models is below the level of the statistical precision.}
  \begin{center}
    \begin{tabular}{llllll}
\hline
 \multicolumn{5}{c}{ Variation in $p_{\rm T}|_{50}$ ($\delta_\text{total}$=  4.9 \%)} \\ \hline 
 $\theta$ & $\delta_{\text{primary}, \, \text{model}}$ (\%) & primary & $\delta_\text{model}$ (\%) & model & $\delta_\text{primary}$ (\%) \\ 
\hline
\multirow{5}{*}{$0^\circ$ } & \multirow{5}{*}{1.95 } &   p &  0.3 & & \\
 & &   Fe &  0.7 & & \\
 & & & & EPOS-LHC &  1.3 \\
 & & & & QGSJetII-04 &  1.4 \\
 & & & & Sibyll 2.3d &  1.6 \\
\hline 
\multirow{5}{*}{$20^\circ$ } & \multirow{5}{*}{2.29 } &   p &  0.7 & & \\
 & &   Fe &  0.9 & & \\
 & & & & EPOS-LHC &  1.2 \\
 & & & & QGSJetII-04 &  1.8 \\
 & & & & Sibyll 2.3d &  1.6 \\
\hline 
\multirow{5}{*}{$40^\circ$ } & \multirow{5}{*}{2.34 } &   p &  0.8 & & \\
 & &   Fe &  0.8 & & \\
 & & & & EPOS-LHC &  1.4 \\
 & & & & QGSJetII-04 &  1.9 \\
 & & & & Sibyll 2.3d &  1.5 \\
\hline 
\multirow{5}{*}{$60^\circ$ } & \multirow{5}{*}{2.35 } &   p &  0.6 & & \\
 & &   Fe &  1.1 & & \\
 & & & & EPOS-LHC &  1.3 \\
 & & & & QGSJetII-04 &  1.8 \\
 & & & & Sibyll 2.3d &  1.8 \\
\hline 
\multirow{5}{*}{$70^\circ$ } & \multirow{5}{*}{2.36 } &   p &  0.7 & & \\
 & &   Fe &  1.0 & & \\
 & & & & EPOS-LHC &  1.4 \\
 & & & & QGSJetII-04 &  2.1 \\
 & & & & Sibyll 2.3d &  1.7 \\
\hline 
\end{tabular}

  \end{center}
  \label{tab:delta-pt}
\end{table}

\begin{table}
  \caption{Variation in the shape of the MPD before the maximum is reached ($X^{\rm L}|_{80}$, see Sect.~\ref{sec:mpd} for the exact definition).}
  \begin{center}
    \begin{tabular}{llllll}
\hline
 \multicolumn{5}{c}{ Variation in $X^{\rm L}|_{80}$ ($\delta_\text{total}$=  8.2 \%)} \\ \hline 
 $\theta$ & $\delta_{\text{primary}, \, \text{model}}$ (\%) & primary & $\delta_\text{model}$ (\%) & model & $\delta_\text{primary}$ (\%) \\ 
\hline
\multirow{5}{*}{$0^\circ$ } & \multirow{5}{*}{6.06 } &   p &  1.3 & & \\
 & &   Fe &  1.7 & & \\
 & & & & EPOS-LHC &  3.8 \\
 & & & & QGSJetII-04 &  4.7 \\
 & & & & Sibyll 2.3d &  4.4 \\
\hline 
\multirow{5}{*}{$20^\circ$ } & \multirow{5}{*}{6.08 } &   p &  1.9 & & \\
 & &   Fe &  1.5 & & \\
 & & & & EPOS-LHC &  4.0 \\
 & & & & QGSJetII-04 &  4.2 \\
 & & & & Sibyll 2.3d &  4.6 \\
\hline 
\multirow{5}{*}{$40^\circ$ } & \multirow{5}{*}{7.43 } &   p &  1.9 & & \\
 & &   Fe &  2.1 & & \\
 & & & & EPOS-LHC &  4.5 \\
 & & & & QGSJetII-04 &  5.5 \\
 & & & & Sibyll 2.3d &  5.3 \\
\hline 
\multirow{5}{*}{$60^\circ$ } & \multirow{5}{*}{6.21 } &   p &  1.6 & & \\
 & &   Fe &  1.9 & & \\
 & & & & EPOS-LHC &  3.5 \\
 & & & & QGSJetII-04 &  4.6 \\
 & & & & Sibyll 2.3d &  4.3 \\
\hline 
\multirow{5}{*}{$70^\circ$ } & \multirow{5}{*}{7.85 } &   p &  2.5 & & \\
 & &   Fe &  2.3 & & \\
 & & & & EPOS-LHC &  4.8 \\
 & & & & QGSJetII-04 &  5.4 \\
 & & & & Sibyll 2.3d &  5.6 \\
\hline 
\end{tabular}

  \end{center}
  \label{tab:delta-x}
\end{table}


\begin{table}
  \caption{Variation in the shower-to-shower fluctuations in the median of the muon energy spectrum.}
  \begin{center}
    \begin{tabular}{llllll}
\hline
 \multicolumn{6}{c}{ Variation in $\sigma_{\rm sh-to-sh}\left [E_{\mu}|_{50}\right ]$ ($\delta_\text{total}$= 253.9 \%)} \\ \hline 
 $\theta$ & $\delta_{\text{primary}, \, \text{model}}$ (\%) & primary & $\delta_\text{model}$ (\%) & model & $\delta_\text{primary}$ (\%) \\ 
\hline
\multirow{5}{*}{$0^\circ$ } & \multirow{5}{*}{155.63 } &   p & 38.7 & & \\
 & &   Fe & 38.9 & & \\
 & & & & EPOS-LHC & 117.5 \\
 & & & & QGSJetII-04 & 102.7 \\
 & & & & Sibyll 2.3d & 123.3 \\
\hline 
\multirow{5}{*}{$20^\circ$ } & \multirow{5}{*}{131.02 } &   p & 44.8 & & \\
 & &   Fe &  7.8 & & \\
 & & & & EPOS-LHC & 80.0 \\
 & & & & QGSJetII-04 & 104.2 \\
 & & & & Sibyll 2.3d & 114.3 \\
\hline 
\multirow{5}{*}{$40^\circ$ } & \multirow{5}{*}{129.67 } &   p & 26.4 & & \\
 & &   Fe & 18.0 & & \\
 & & & & EPOS-LHC & 103.8 \\
 & & & & QGSJetII-04 & 107.1 \\
 & & & & Sibyll 2.3d & 110.0 \\
\hline 
\multirow{5}{*}{$60^\circ$ } & \multirow{5}{*}{161.18 } &   p & 56.0 & & \\
 & &   Fe & 45.6 & & \\
 & & & & EPOS-LHC & 106.7 \\
 & & & & QGSJetII-04 & 95.9 \\
 & & & & Sibyll 2.3d & 112.1 \\
\hline 
\multirow{5}{*}{$70^\circ$ } & \multirow{5}{*}{132.94 } &   p & 40.2 & & \\
 & &   Fe & 33.3 & & \\
 & & & & EPOS-LHC & 92.6 \\
 & & & & QGSJetII-04 & 91.4 \\
 & & & & Sibyll 2.3d & 111.0 \\
\hline 
\end{tabular}

  \end{center}
  \label{tab:delta-sh2sh-en}
\end{table}

\begin{table}
  \caption{Variation in the shower-to-shower fluctuations in the median of the transverse momentum spectrum.}
  \begin{center}
    \begin{tabular}{llllll}
\hline
 \multicolumn{6}{c}{ Variation in $\sigma_{\rm sh-to-sh}\left [p_{\rm T}|_{50}\right ]$ ($\delta_\text{total}$= 130.7 \%)} \\ \hline 
 $\theta$ & $\delta_{\text{primary}, \, \text{model}}$ (\%) & primary & $\delta_\text{model}$ (\%) & model & $\delta_\text{primary}$ (\%) \\ 
\hline
\multirow{5}{*}{$0^\circ$ } & \multirow{5}{*}{103.78 } &   p & 12.7 & & \\
 & &   Fe & 19.0 & & \\
 & & & & EPOS-LHC & 85.9 \\
 & & & & QGSJetII-04 & 82.7 \\
 & & & & Sibyll 2.3d & 101.7 \\
\hline 
\multirow{5}{*}{$20^\circ$ } & \multirow{5}{*}{98.61 } &   p & 25.1 & & \\
 & &   Fe & 23.0 & & \\
 & & & & EPOS-LHC & 80.2 \\
 & & & & QGSJetII-04 & 61.8 \\
 & & & & Sibyll 2.3d & 74.1 \\
\hline 
\multirow{5}{*}{$40^\circ$ } & \multirow{5}{*}{113.58 } &   p & 46.5 & & \\
 & &   Fe & 14.3 & & \\
 & & & & EPOS-LHC & 61.4 \\
 & & & & QGSJetII-04 & 94.7 \\
 & & & & Sibyll 2.3d & 92.3 \\
\hline 
\multirow{5}{*}{$60^\circ$ } & \multirow{5}{*}{108.71 } &   p & 28.8 & & \\
 & &   Fe &  2.3 & & \\
 & & & & EPOS-LHC & 73.1 \\
 & & & & QGSJetII-04 & 85.5 \\
 & & & & Sibyll 2.3d & 96.9 \\
\hline 
\multirow{5}{*}{$70^\circ$ } & \multirow{5}{*}{99.42 } &   p & 18.7 & & \\
 & &   Fe & 33.0 & & \\
 & & & & EPOS-LHC & 84.3 \\
 & & & & QGSJetII-04 & 65.2 \\
 & & & & Sibyll 2.3d & 79.1 \\
\hline 
\end{tabular}

  \end{center}
  \label{tab:delta-sh2sh-pt}
\end{table}

\begin{table}
  \caption{Variation in the shower-to-shower fluctuations of the shape of the MPD before the maximum is reached ($X^{\rm L}|_{80}$, see Sect.~\ref{sec:mpd} for the exact definition).}
  \begin{center}
    \begin{tabular}{llllll}
\hline
 \multicolumn{6}{c}{ Variation in $\sigma_{\rm sh-to-sh}\left [X^{\rm L}|_{80}\right ]$ ($\delta_\text{total}$= 140.0 \%)} \\ \hline 
 $\theta$ & $\delta_{\text{primary}, \, \text{model}}$ (\%) & primary & $\delta_\text{model}$ (\%) & model & $\delta_\text{primary}$ (\%) \\ 
\hline
\multirow{5}{*}{$0^\circ$ } & \multirow{5}{*}{86.09 } &   p & 16.1 & & \\
 & &   Fe & 21.7 & & \\
 & & & & EPOS-LHC & 72.8 \\
 & & & & QGSJetII-04 & 67.7 \\
 & & & & Sibyll 2.3d & 69.7 \\
\hline 
\multirow{5}{*}{$20^\circ$ } & \multirow{5}{*}{105.12 } &   p & 52.7 & & \\
 & &   Fe & 21.6 & & \\
 & & & & EPOS-LHC & 45.2 \\
 & & & & QGSJetII-04 & 58.3 \\
 & & & & Sibyll 2.3d & 84.4 \\
\hline 
\multirow{5}{*}{$40^\circ$ } & \multirow{5}{*}{112.18 } &   p & 36.7 & & \\
 & &   Fe & 10.1 & & \\
 & & & & EPOS-LHC & 71.2 \\
 & & & & QGSJetII-04 & 84.7 \\
 & & & & Sibyll 2.3d & 96.8 \\
\hline 
\multirow{5}{*}{$60^\circ$ } & \multirow{5}{*}{104.74 } &   p & 27.9 & & \\
 & &   Fe & 30.4 & & \\
 & & & & EPOS-LHC & 82.6 \\
 & & & & QGSJetII-04 & 59.1 \\
 & & & & Sibyll 2.3d & 73.7 \\
\hline 
\multirow{5}{*}{$70^\circ$ } & \multirow{5}{*}{124.03 } &   p & 47.4 & & \\
 & &   Fe & 29.4 & & \\
 & & & & EPOS-LHC & 74.4 \\
 & & & & QGSJetII-04 & 71.9 \\
 & & & & Sibyll 2.3d & 88.8 \\
\hline 
\end{tabular}

  \end{center}
  \label{tab:delta-sh2sh-x}
\end{table}

\end{document}